\documentclass[aps,prb,twocolumn,showpacs,citeautoscript,reprint,longbibliography,superscriptaddress]{revtex4-2}
\usepackage{graphicx}
\usepackage{bm}
\usepackage{natbib}
\usepackage{amsmath}
\usepackage{xfrac}
\usepackage{nicefrac}
\usepackage[colorlinks=true, urlcolor=blue, linkcolor=blue, citecolor=blue, pdfborder={0 0 0}]{hyperref}
\usepackage[caption=false]{subfig}
\usepackage{cleveref}
\usepackage{dcolumn}

\crefname{figure}{Fig.}{Figs}
\crefname{table}{Table}{Tables}
\crefname{section}{Sec.}{Secs.}

\begin{document}

\title{Calculating interface transport parameters at finite temperatures: \\
nonmagnetic interfaces}
	
\author{Kriti Gupta}
\affiliation{Faculty of Science and Technology and MESA$^+$ Institute for Nanotechnology, University of Twente, P.O. Box 217,
	7500 AE Enschede, The Netherlands}
\author{Ruixi Liu}
\affiliation{Center for Advanced Quantum Studies and Department of Physics, Beijing Normal University, 100875 Beijing, China}
\author{Rien J. H. Wesselink}
\affiliation{Faculty of Science and Technology and MESA$^+$ Institute for Nanotechnology, University of Twente, P.O. Box 217, 7500 AE Enschede, The Netherlands}
\author{Zhe Yuan}
\email[Email: ]{zyuan@bnu.edu.cn}
\affiliation{Center for Advanced Quantum Studies and Department of Physics, Beijing Normal University, 100875 Beijing, China}
\author{Paul J. Kelly}
\email[Email: ]{P.J.Kelly@utwente.nl}
\affiliation{Faculty of Science and Technology and MESA$^+$ Institute for Nanotechnology, University of Twente, P.O. Box 217,
	7500 AE Enschede, The Netherlands}
\affiliation{Center for Advanced Quantum Studies and Department of Physics, Beijing Normal University, 100875 Beijing, China}
	
\date{\today}
		
\begin{abstract}
First-principles scattering calculations are used to investigate spin transport through interfaces between diffusive nonmagnetic metals where the symmetry lowering leads to an enhancement of the effect of spin-orbit coupling (SOC) and to a discontinuity of the spin currents passing through the interfaces. From the conductance and local spin currents calculated for nonmagnetic bilayers, we extract values of the room temperature interface resistance $R_{\rm I}$, of the spin memory loss parameter $\delta$ and of the interface spin Hall angle $\Theta_{\rm I}$ for nonmagnetic Au$|$Pt and Au$|$Pd interfaces using a frozen thermal disorder scheme to model finite temperatures. Substantial values of all three parameters are found with important consequences for experiments involving nonmagnetic spacer and capping layers. The temperature dependence of the interface parameters is determined for Au$|$Pt.
\end{abstract}
	
	
\maketitle

\section{Introduction}
\label{sec:Introduction}

With the discovery of the giant magnetoresistance (GMR) effect in magnetic multilayers \cite{Baibich:prl88, Binasch:prb89}, it was recognized that interfaces play a key role in spin transport phenomena. In semiclassical formulations of transport \cite{vanSon:prl87, Valet:prb93, Gijs:ap97, Brataas:prp06, Bass:jmmm16} they appear as discrete resistances and the description of the transport of a current of electrons through a nonmagnetic NM$|$NM$'$ bilayer comprising two nonmagnetic metals then requires three parameters: two resistivities $\rho$ and $\rho'$ and the interface resistance $R_{\rm I}$. Because spin is not conserved when SOC is included, describing its transport requires introducing a spin-flip diffusion length (SDL) in each material, $l_{\rm sf}$ and $l'_{\rm sf}$ as well as its interface counterpart, the spin memory loss (SML) parameter $\delta$. Thus, to describe spin transport through a NM$|$NM$'$ bilayer requires a total of six parameters. While bulk resistivities are readily measured, determining $l_{\rm sf}$ remains controversial; for well studied materials like Pt, values reported over the last decade span an order of magnitude \cite{Bass:jpcm07, Sinova:rmp15, Wesselink:prb19}. Almost everything we know about interface parameters is from current-perpendicular-to-the-plane (CPP) magnetoresistance experiments \cite{Bass:jpcm07, Bass:jmmm16} interpreted using the semiclassical Valet-Fert (VF) model \cite{Valet:prb93}.  While these experiments are relatively simple to interpret, they are restricted to low temperatures as they require superconducting leads \cite{Bass:jmmm16}. Because the vast majority of experimental studies in spintronics is carried out at room temperature, there is a need to know how transport parameters, in particular those describing interfaces, behave as a function of temperature. 

This need is accentuated by the huge interest in recent years \cite{Hoffmann:ieeem13, Sinova:rmp15} in the spin Hall effect (SHE) \cite{Dyakonov:zetf71, *Dyakonov:pla71, Hirsch:prl99, Zhang:prl00} whereby a longitudinal charge current excites a transverse spin current in nonmagnetic materials, and in its inverse, the inverse SHE (ISHE). Determination of the spin Hall angle (SHA) $\Theta_{\rm sH}$ that measures the efficiency of the SHE is intimately connected with the spin-flip diffusion length and, because an interface is always involved, with the SML \cite{Rojas-Sanchez:prl14}. When use is made of spin pumping and the ISHE \cite{Saitoh:apl06, Ando:prl08, Mosendz:prl10, Mosendz:prb10} or the SHE and spin-transfer torque (STT) \cite{Liu:prl11}, the interface in question is an FM$|$NM interface between ferromagnetic and nonmagnetic materials. When the nonlocal spin injection method is used \cite{Kimura:prl07, Vila:prl07}, two interfaces are involved: an FM$|$NM interface to create a spin accumulation and an NM$|$NM$'$ interface to detect it. Recent studies suggest that measurements of the SHA may actually be dominated by interface effects \cite{Rojas-Sanchez:prl14, LiuY:prl14, WangL:prl16, Amin:prb16a} and that the experimental determination of interface and bulk parameters are inextricably coupled. 

In such a situation, it is crucial to have a way of determining the interface parameters independently. We recently described a formalism to evaluate local charge and spin currents \cite{Wesselink:prb19} from the solutions of fully relativistic quantum mechanical scattering calculations \cite{Starikov:prl10, *Starikov:prb18} that include temperature-induced lattice and spin disorder \cite{LiuY:prb11, LiuY:prb15}. This yielded a layer-resolved description of spin currents propagating through atomic layers of thermally disordered Pt and Py, that allowed us to unambiguously determine bulk transport properties. For Pt, large deviations from bulk behaviour were observed close to the interfaces with the (ballistic) Au leads that pointed towards strong interface SOC effects. In this paper we focus on interface transport properties and study realistic interfaces between thermally disordered materials. By focussing on charge and spin currents rather than scattering matrices \cite{Belashchenko:prl16, *Flores:prb20, Dolui:prb17}, we will  evaluate the parameters entering the VF semiclassical formalism \cite{Valet:prb93} that is universally used to interpret current-perpendicular-to-the-plane (CPP) experiments \cite{Bass:jmmm16} as well as the interface SHA. 
Since everything we know about nonmagnetic interfaces is through low temperature magnetoresistance experiments \cite{Galinon:apl05, Bass:jmmm16} or through calculations for ballistic interfaces \cite{Schep:prb97, Stiles:prb00, Xia:prb01, Bauer:jpd02, Xu:prl06, Belashchenko:prl16, *Flores:prb20, Dolui:prb17}, we will also investigate the temperature dependence of interface transport parameters.

\begin{figure}[t]
\centering
\includegraphics[width=8.6cm]{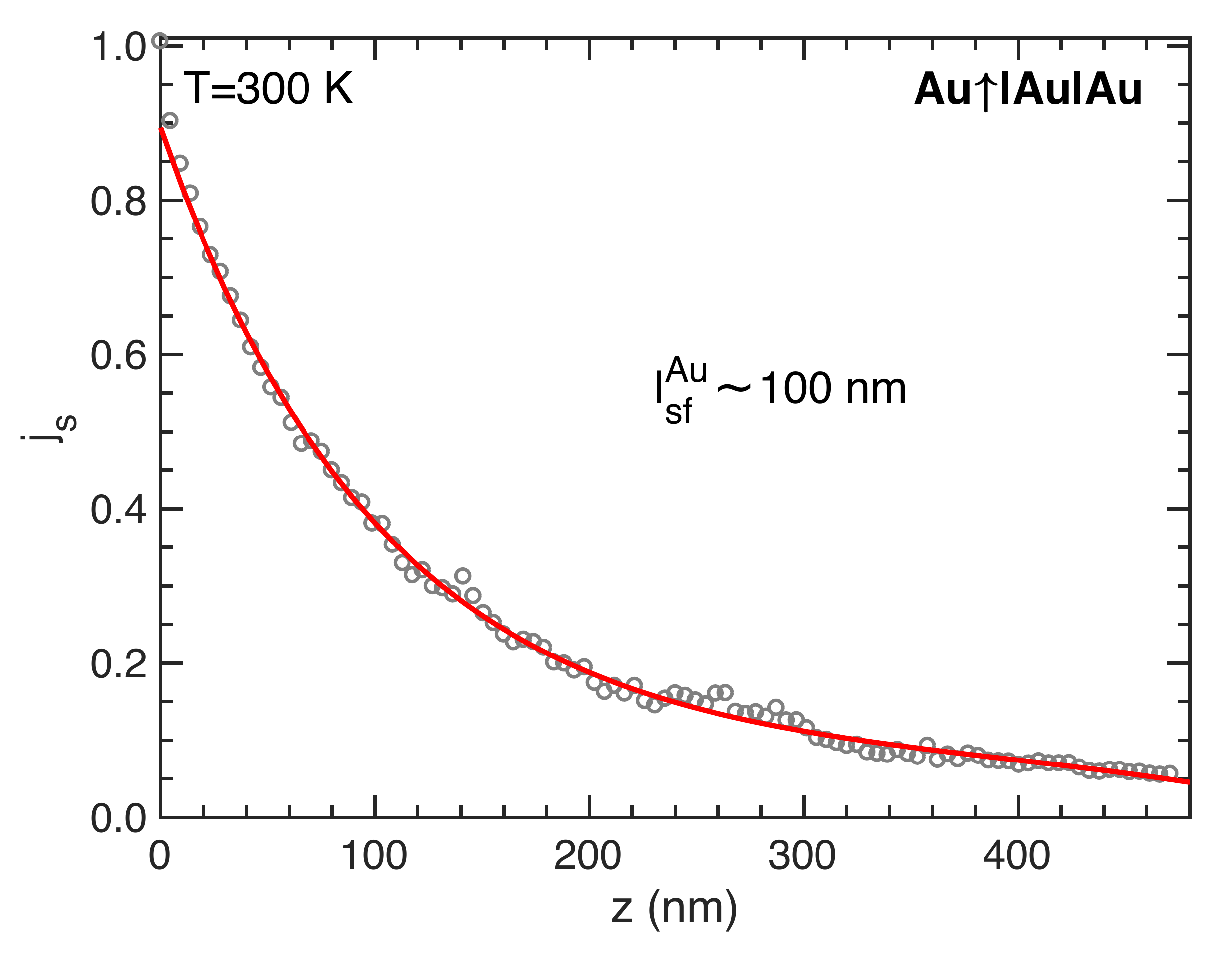}
\caption{A fully spin-polarized current [$j_s(0)=1$] is injected into 470 nm of diffusive Au at 300 K for a $4 \times 4$ lateral supercell. The open circles indicate the calculated spin current profile. An exponential fit yields $l_{\rm Au} \sim 100$ nm. }
\label{fig1}
\end{figure}

The NM$|$NM$'$ interfaces we will investigate are Au$|$Pd and Au$|$Pt. This choice offers several advantages. Au has a low effective SOC because its $d$ and $p$ bands are completely filled or empty and only affect transport via hybridization with the free-electron-like $s$ band that is Kramers degenerate for an inversion symmetric material. This is expected to lead to a very long spin-flip diffusion length and small SHA. At room temperature, literature values of $l_{\rm Au} \equiv l_{\rm sf}^{\rm Au}$ are reported in the range 25-86~nm \cite{Sinova:rmp15} while $\Theta_{\rm Au} \equiv \Theta_{\rm sH}^{\rm Au}$ is found to be as small as 0.05\% \cite{Isasa:prb15a} or as large as 11.3\% \cite{Seki:natm08, Sinova:rmp15} and apparently depends strongly on the thickness of the samples used in the measurements \cite{Seki:natm08, Seki:ssc10, Gu:prl10}. We can estimate $l_{\rm Au}$ at room temperature by injecting a fully spin-polarized current from a ``half-metallic ferromagnetic'' Au lead (Au$\uparrow$) \cite{Starikov:prb18, Wesselink:prb19} into a long scattering region composed of diffusive Au. The results obtained for a single configuration of thermal disorder, a small $4 \times 4$ lateral supercell and a 470 nm thick slab of diffusive Au are plotted in \cref{fig1}. By fitting the exponentially decaying spin current, that does not fully decrease to zero, we obtain an estimate  of $l_{\rm Au} \sim 100$ nm. A more detailed study \footnote{The present estimate is not converged with respect to the lateral supercell size and is only a rough estimate. A better estimate of 50.9 nm can be obtained from a systematic study at an elevated temperature of 1000~K \cite{Nair:tbp22}. This makes it possible to determine the dependence on the lateral supercell size for scattering regions that are both sufficiently long and computationally tractable and extrapolate to infinite lateral supercell size. By focussing on the product $\rho(T) l_{\rm sf}(T)$, that according to the Elliott-Yafet relationship is independent of temperature \cite{Nair:prl21}, the room temperature value can be obtained as $l_{\rm sf}(300)=\rho(1000) l_{\rm sf}(1000)/\rho(300)$ } yields a better estimate of 50.9 nm \cite{Nair:prl21}.
In order to determine the interface spin memory loss, we will use this Au$\uparrow$$|$Au construction to inject a fully spin-polarized current into a thin slab of diffusive Au that will undergo minimal decay before encountering an interface with either Pt or Pd. The lattice mismatch between Au, Pt and Pd is small enough to construct pseudomorphic interfaces by compressing Au uniformly to match the other two lattices without drastically changing the electronic structure of Au at the Fermi energy. The effect of this approximation will be examined by constructing interfaces between fully relaxed lattices. 

Besides evaluating $\delta$, we will look to see if there is an interface spin Hall effect at the Au$|$Pt and Au$|$Pd interfaces. Since the prediction of Wang {\em et al.} of such an effect for Py$|$Pt \cite{WangL:prl16}, there have been experiments \cite{Jungfleisch:prb16} and theoretical studies \cite{Amin:prb16a, *Amin:prb16b, Amin:prl18} that point towards a role for interface SOC in generating spin currents at NM$|$NM$'$ interfaces. 

The plan of the paper is as follows. In \Cref{sec:Methods3} we first summarize the original VF model (\cref{subsec:VF}) and then extend it to include the effect of SOC at interfaces and discuss how it will be used to extract interface parameters (\cref{subsec:intdis}). We describe how transverse spin currents generated by the spin Hall effect behave at interfaces (\cref{subsec:shaint}) before elaborating on the scheme we use to extract quantitative estimates of the interface SHA from ISHE calculations for interfaces (\cref{subsec:ishaint}); a more general scheme is given in an appendix. 
In \cref{sec:Calculations3} we briefly summarize a number of important features of our first-principles scattering theory \cite{Xia:prb06, Starikov:prb18, Wesselink:prb19}, give details of how Au$|$Pt and Au$|$Pd interface geometries are constructed and describe how temperature is incorporated in the adiabatic approximation.
In \cref{sec:Results3}, we demonstrate the procedure described in \cref{sec:Methods3} by extracting the interface parameters for Au$|$Pt at room temperature: interface resistance (Sec. \ref{subsec:intRAuPt}), spin memory loss (Sec. \ref{subsec:delta}), interface spin Hall angle (Secs. \ref{subsec:sha} and \ref{subsec:thetaI}), compare them with interface parameters for Au$|$Pd (\cref{subsec:AuPt_AuPd}) and determine Au$|$Pt parameters at 200 and 400 K (\cref{subsec:temp}). In \cref{sec:Conclusions} we summarize our findings and present some conclusions. Details of a parallel study of interfaces between nonmagnetic and ferromagnetic materials can be found in \cite{Gupta:prb21} and a brief report of both appeared in Ref.~\cite{Gupta:prl20}.

\section{Methods}
\label{sec:Methods3}

\subsection{Valet-Fert model}
\label{subsec:VF}

In this subsection, we recapitulate the VF model \cite{Valet:prb93} for spin transport before generalizing it to include spin-flip scattering at interfaces \cite{Baxter:jap99}. The macroscopic equations derived by Valet and Fert characterize transport in terms of material-specific parameters. For a current flowing along the $z$ direction perpendicular to the interface plane in an axially symmetric CPP geometry, the spatial profiles of majority (minority) spin current densities $j_{\uparrow(\downarrow)}$ and chemical potentials $\mu_{\uparrow(\downarrow)}$ are related as follows 
\begin{subequations}
\begin{eqnarray}
\frac{\partial^2 \mu_s}{\partial z^2}&=&\frac{\mu_s}{l^2}, \label{eq:diffusion}\\
j_{\uparrow(\downarrow)}(z)&=&-\frac{1}{e \rho_{\uparrow(\downarrow)}}\frac{\partial\mu_{\uparrow(\downarrow)}}{\partial z}. \label{eq:ohm}
\end{eqnarray}
\end{subequations}
Here, $\mu_s=\mu_\uparrow-\mu_\downarrow$ is the spin accumulation, $l\equiv l_{\rm sf}$ denotes the spin-flip diffusion length, and $\rho_{\uparrow(\downarrow)}$ is a spin-dependent bulk resistivity. According to the ``two-current series-resistor'' (2CSR) model \cite{Zhang:jap91, Lee:jmmm93, Valet:prb93}, resistances are first calculated separately for spin-up and spin-down electrons and then added in parallel. For non-magnetic materials, $\rho_{\uparrow}=\rho_{\downarrow}=2\rho$, where
$\rho$ is the total resistivity. Thus, spin transport in the bulk of a  material can be characterized in terms of its resistivity $\rho$ and spin-flip diffusion length $l$. Equations (\ref{eq:diffusion}) and (\ref{eq:ohm}) can be solved for $\mu_\uparrow$,  $\mu_\downarrow$, $j_\uparrow$, $j_\downarrow$ making use of the condition that the total current density $j=j_\uparrow+j_\downarrow$ is conserved in one-dimensional transport. The general solution of \eqref{eq:diffusion} is $\mu_s(z)=A e^{z/l}+B e^{-z/l}$. The normalized spin-current density $j_s(z) \equiv [j_\uparrow(z)-j_\downarrow(z)]/j$ is given by 
\begin{equation}
j_s(z) = \frac{1}{2ej\rho l}\Big[B e^{-z/l}-A e^{z/l}\Big]\label{eq:js3}
\end{equation}
where the coefficients $A$ and $B$ are to be determined by using suitable boundary conditions. 

Spin-dependent scattering at an interface is characterized in terms of the interface resistances $R_{\uparrow(\downarrow)}$. By analogy with the bulk resistivity, $R_{\uparrow(\downarrow)}=2R_{\rm I}$ where $R_{\rm I}$ is the total interface resistance.

\subsection{Interface discontinuity}
\label{subsec:intdis}

\begin{figure}[t]
\includegraphics[width=8.2cm]{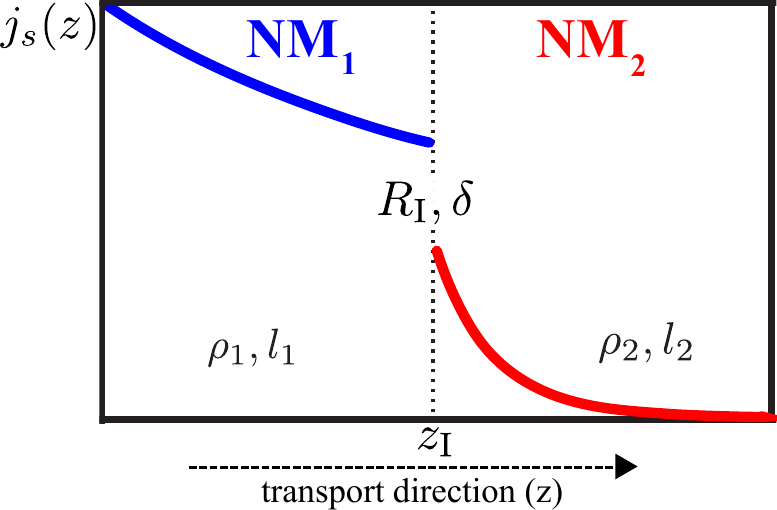}
\caption{Generalized VF model for spin-flip scattering in a nonmagnetic bilayer. A fully spin-polarized current is injected into ${\rm NM_1}$ from a ballistic half-metallic lead. The decay of current in NM$_i$ is parametrized by the resistivity $\rho_i$ and spin-flip diffusion length $l_i$ of material $i=1,2$. Because of the enhancment of the effective spin-orbit interaction at an interface, the spin current is discontinuous at $z_{\rm I}$ (position of the geometrical interface). This discontinuity is characterized in terms of the interface resistance $R_{\rm I}$ and spin-memory loss $\delta$.}
\label{fig2}
\end{figure}

The  model described above was extended by Fert and Lee \cite{Fert:prb96b} to include the effect of interface SOC in terms of additional spin-flip interface resistances. The effect of interface spin flipping  was first described in terms of the spin-memory loss parameter $\delta$ by Baxter {\em et al.} for NM$|$NM$'$ interfaces between two nonmagnetic metals \cite{Baxter:jap99}. In this subsection, we summarize this generalized VF model and extract the boundary conditions for a geometrically sharp NM$|$NM$'$ interface.

In \cref{fig2} we sketch how a spin current passes through a nonmagnetic bilayer. A fully spin-polarized current $j_s(0)=1$ enters the first diffusive nonmagnetic layer, NM$_1$, from the left. The SDL in this material, $l_1$, determines the exponential decay of $j_s(z)$. An interface breaks inversion symmetry and lifts the Kramers degeneracy. The effect of SOC-induced energy band splittings is enhanced by the symmetry lowering so that the spin current decays rapidly in the vicinity of the interface leading to an interface discontinuity in $j_s(z)$. In the semiclassical framework, the discontinuity is quantified in terms of the spin memory loss parameter $\delta$. After the abrupt decay at the interface, the spin current that survives in ${\rm NM_2}$ decays to zero on a length scale described by the SDL $l_2$. 

\begin{figure}[t]
\includegraphics[width=8.6cm]{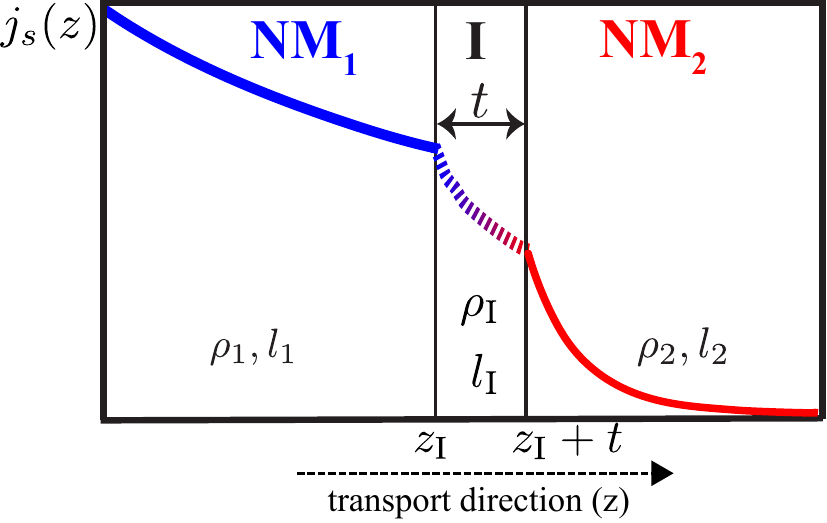}
\caption{VF model for a ${\rm NM_1|I|NM_2}$ trilayer. The interface is modelled as a fictitious bulk-like layer with thickness $t$ that is characterized by parameters $\rho_{\rm I}$ and $l_{\rm I}$, such that $\rho_{\rm I}=AR_{\rm I}/t$ and $l_{\rm I}=t/\delta$. This spin current is continuous at the ${\rm NM_1|I}$ and ${\rm I|NM_2}$ interfaces.}
\label{fig3}
\end{figure}

By fitting $j_s(z)$ calculated quantum mechanically from scattering theory to the corresponding VF equation, we obtain values of $j_s(z_{\rm I})$ on either side of the interface, $z=z_{\rm  I}\pm \epsilon$. To extract $\delta$, the interface discontinuity is incorporated into the VF framework. It is assumed that the interface region (I) has a finite thickness $t$ and can be treated as a material with resistivity $\rho_{\rm I}$ and spin-flip diffusion length $l_{\rm I}$. In terms of these ``bulk'' parameters, the areal interface resistance $AR_{\rm I}$ and SML $\delta$ are
\begin{equation}
AR_{\rm I} = \lim_{t \to 0} \rho_{\rm I} t  \,\,\,\,\,\,{\rm and} \,\,\,\,\,\, \delta = \lim_{t \to 0} t/l_{\rm I} .
\label{eq:intpar}
\end{equation}
With the above description of the interface, a bilayer of any two non-magnetic bulk materials (${\rm NM_1}$ and ${\rm NM_2}$) becomes a fictitious trilayer NM$_1|$I$|$NM$_2$. Spin transport in this geometry can thus be characterized by six bulk transport parameters $\rho_1,~\rho_{\rm I},~\rho_2,~l_1,~l_{\rm I},~l_2$ instead of $\rho_1,~\rho_2,~AR_{\rm I},~l_1,~l_2$ and $\delta$. The generalized spin transport equations for the three distinct layers labelled $i=1, 2$ and I are 
\begin{subequations}
\begin{eqnarray}
\mu_{si}(z)&=&A_ie^{z/l_i}+B_i e^{-z/l_i},\\
  j_{si}(z)&=&\frac{1}{2ej\rho_i l_i}\Big[B_i e^{-z/l_i}-A_i e^{z/l_i}\Big] 
\label{eq:jsi3}.
\end{eqnarray}
\end{subequations}

To switch from an ${\rm NM_1}|{\rm NM_2}$ to an ${\rm NM_1}|$I$|{\rm NM_2}$ description, the discontinuity at the sharp interface at $z=z_{\rm I}$ in $\mu_s(z_{\rm I})$ and $j_s(z_{\rm I})$ in \cref{fig2} becomes a continuous transition through the interface layer between $z=z_{\rm I}$ and $z=z_{\rm I}+t$ in \cref{fig3}. Continuity of $\mu(z)$ and $j_s(z)$ at the ${\rm NM_1|I}$ and ${\rm I|NM_2}$ interfaces yields the equations 
\begin{subequations}
\begin{eqnarray}
\mu_{s1}(z_{\rm I})  &=&A_{\rm I} e^{z_{\rm I}/l_{\rm I}}+B_{\rm I} e^{-z_{\rm I}/l_{\rm I}}, \\
\mu_{s2}(z_{\rm I}+t)&=&A_{\rm I} e^{z_{\rm I}/l_{\rm I}}e^{\delta}+B_{\rm I} e^{-z_{\rm I}/l_{\rm I}}e^{-\delta},
\end{eqnarray}
\end{subequations}
and
\begin{subequations}
\begin{eqnarray}
\!\! j_{s1}(z_{\rm I})&=&\frac{1}{2ej\rho_{\rm I} l_{\rm I}}
\Big[B_{\rm I} e^{-z_{\rm I}/l_{\rm I}}-A_{\rm I} e^{z_{\rm I}/l_{\rm I}}\Big],\\
\!\! j_{s2}(z_{\rm I}+t)&=&\frac{1}{2ej\rho_{\rm I} l_{\rm I}}
\Big[B_{\rm I} e^{-z_{\rm I}/l_{\rm I}}e^{-\delta}-A_{\rm I} e^{z_{\rm I}/l_{\rm I}}e^{\delta}\Big].
\end{eqnarray} 
\end{subequations}
Eliminating $A_{\rm I}$ and $B_{\rm I}$ and taking the limit $t\rightarrow0$ leads to the expected discontinuity in $\mu_s$ and $j_s$ at the NM$_1|{\rm NM_2}$ interface. Substituting \eqref{eq:intpar} then yields 
\begin{subequations}
\begin{eqnarray}
\!\!\!\!\!\! j_{s1}(z_{\rm I})&=&\frac{\delta}{2ejAR_{\rm I} \sinh{\delta}}
          \Big[\mu_{s1}(z_{\rm I})\cosh{\delta}-\mu_{s2}(z_{\rm I}) \Big],\\
\!\!\!\!\!\! j_{s2}(z_{\rm I})&=&\frac{\delta}{2ejAR_{\rm I} \sinh{\delta}}
          \Big[\mu_{s1}(z_{\rm I})-\mu_{s2}(z_{\rm I})\cosh{\delta} \Big].
\end{eqnarray}
\label{eq:intmueq}
\end{subequations}
Use of the remaining boundary conditions: $j_s(0)=1, j_s(\infty)=0$ allows us to express $\mu_{s1}(z_{\rm I})$ and $\mu_{s2}(z_{\rm I})$ in terms of $j_{s1}(z_{\rm I})$ and $j_{s2}(z_{\rm I})$. After some algebra, we obtain
\begin{widetext}
\begin{subequations}
\label{eq:jsI}
\begin{align}
j_{s1}(z_{\rm I})=\frac{\delta}{R_{\rm I} \sinh\delta}
\Bigg[\rho_1l_1 \cosh\delta  
\Big\{\coth\Big(\frac{z_{\rm I}}{l_1}\Big) \big[j_{s1}(z_{\rm I})-e^{z_{\rm I}/l_1}\big]+e^{z_{\rm I}/l_1}\Big\}-\rho_2 l_2 \, j_{s2}(z_{\rm I})\Bigg],\\
j_{s2}(z_{\rm I})=\frac{\delta}{R_{\rm I} \sinh\delta}
\Bigg[\rho_1l_1 
\Big\{\coth\Big(\frac{z_{\rm I}}{l_1}\Big)\big[j_{s1}(z_{\rm I})-e^{z_{\rm I}/l_1}\big]+e^{z_{\rm I}/l_1}\Big\}-\rho_2 l_2 \, j_{s2}(z_{\rm I}) \cosh\delta \Bigg].
		\end{align}
\end{subequations}
\end{widetext}	

In principle we can solve either of the above two equations numerically by substituting values of all parameters and $j_{s1}(z_{\rm I})$ and $j_{s2}(z_{\rm I})$ to find $\delta$. However, as mentioned in the introduction, $l_{\rm Au}$ cannot be easily determined accurately. To avoid using $l_{\rm Au}$ in extracting $\delta$, we eliminate $\rho_1l_1$ in the above two equations to yield
\begin{equation}
\label{eq:delta}
\frac{j_{s1}(z_{\rm I})}{j_{s2}(z_{\rm I})}=\cosh{\delta} + \delta \sinh{\delta}
\,\, \frac{\rho_2 l_2}{AR_{\rm I}} 
\end{equation}
expressing $\delta$ in terms of $j_{s1}(z_{\rm I})$ and $j_{s2}(z_{\rm I})$ as well as $\rho_2,~l_2$ and $R_{\rm I}$. 

$\rho_2$ and $AR_{\rm I}$ can be determined from calculations of the conductance using the Landauer-B{\"{u}}ttiker relationship. The spin-flip diffusion length $l_2$ is extracted using the method illustrated in \cref{fig1} for Au and described in more detail in \cite{Wesselink:prb19}. $j_{s1}(z_{\rm I})$ and $j_{s2}(z_{\rm I})$ will be determined by fitting the spin current in $\rm NM_1|NM_2$ using \eqref{eq:jsi3}.

\begin{figure}[!b]
\centering
\includegraphics[width=8.6cm]{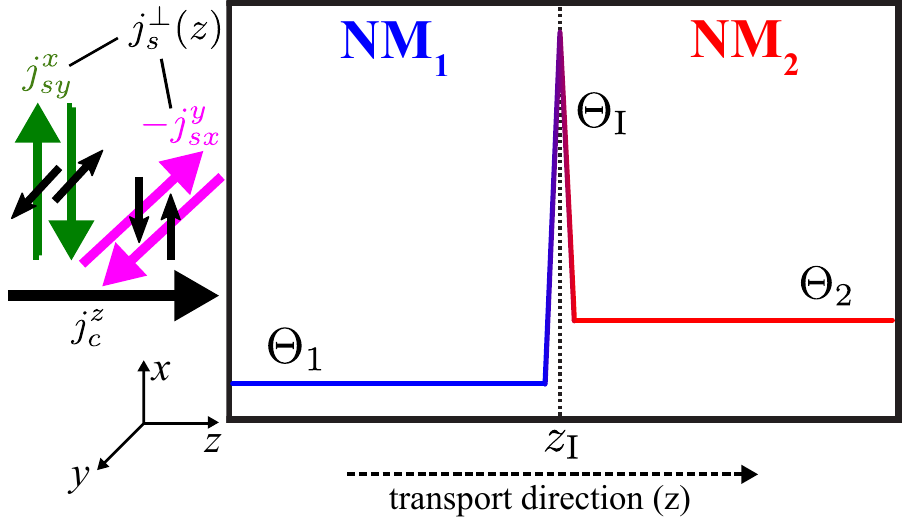}
\caption{Sketch of the transverse spin currents $j_{s\alpha}^{\perp}$ ($j_{sy}^x$, vertical green arrows and $-j_{sx}^y$, pink arrows) generated in response to a constant charge current $j_c^z=j$ (horizontal black arrow) through an ${\rm NM_1}|{\rm NM_2}$ bilayer. $\Theta_1$ and $\Theta_2$ represent the ratio $j_{s\alpha}^{\perp}/j$ in ${\rm NM_1}$ and ${\rm NM_2}$ respectively. A sharp peak in $j_s^{\perp}$ at the interface is attributed to an interface spin Hall effect described by the angle $\Theta_{\rm I}$.}
	\label{fig4}
\end{figure}

\subsection{Transverse spin current at an interface}
\label{subsec:shaint}

When a charge current is passed through a nonmagnetic bulk material, the SOC leads to a transverse spin current in an effect called the spin Hall effect \cite{Dyakonov:zetf71, *Dyakonov:pla71, Hirsch:prl99, Hoffmann:ieeem13, Sinova:rmp15}. This spin current can be denoted $j_{s\alpha}^{\perp}$ where $\alpha$ labels the direction of the spin polarization that is given by the vector product of the driving charge current and the induced transverse spin current. As sketched in \cref{fig4} for a constant charge current in the $z$ direction, $j\equiv j_c^z$, transverse spin currents are generated in the radial direction e.g. $x$ and $-y$ directions that are polarized in the $y$ and $x$ directions, respectively. The amount of spin current generated per unit charge current is given by the SHA $\Theta_{\rm sH}=j_{s\alpha}^{\perp}/j$. By measuring charge currents in terms of the fundamental unit of charge $-|e|$ and spin currents in units of $\hbar/2$, $\Theta_{\rm sH}$ becomes dimensionless.

When the constant charge current $j$ passes through a bilayer composed of the nonmagnetic materials NM$_1$ and NM$_2$ perpendicular to the NM$_1|$NM$_2$ interface, it  gives rise to transverse spin currents $j_{s\alpha}^{\perp}$ with magnitudes given by $j \Theta_1$ in NM$_1$ and $j \Theta_2$ in NM$_2$, respectively; see  \cref{fig4}. In the vicinity of the interface, the possibility of an abrupt deviation from the bulk behaviour resulting from an interface spin-Hall effect and described by the angle $\Theta_{\rm I}$ has been proposed \cite{WangL:prl16}. Because of the finite width of this peak, it is not possible to directly extract $\Theta_{\rm I}$ from the transverse spin currents. 
Instead, we will follow Wang {\em et al.} \cite{WangL:prl16} and access it through the ISHE whereby a spin current polarized perpendicular to the current direction generates a transverse charge current along a mutually perpendicular direction. How $\Theta_{\rm I}$ can be extracted from ISHE calculations is described in the following subsection.

\subsection{Transverse charge current and interface ISHE}
\label{subsec:ishaint}

\begin{figure}[t]
\includegraphics[width=8.6cm]{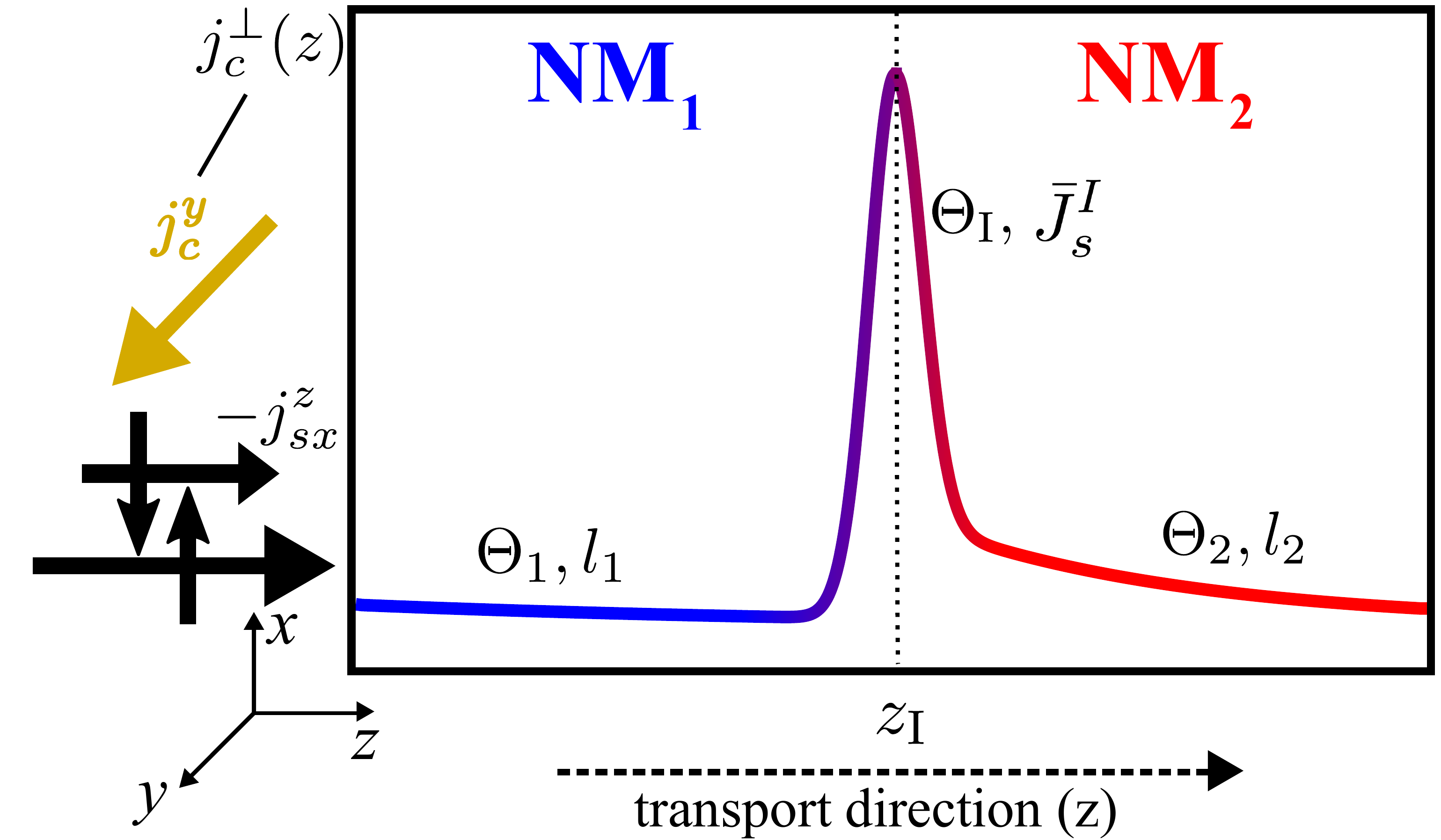}
\caption{On injecting a spin polarized current, whose polarization is perpendicular to the current direction, into a NM$_1|$NM$_2$ bilayer, a transverse charge current $j_c^{\perp}$ is generated in a mutually perpendicular direction because of the ISHE. The spin current $j_{sx}^z(z)$, composed of up and down spins propagating to the right (black horizontal arrows), is not conserved because of SOC-induced spin flipping. It induces a transverse charge current in each layer determined by the respective SHAs and results in the spatially varying $j_c^{\perp}(z)$ sketched in the figure. At the interface, the discontinuity in $j_s(z)$ that is spin memory loss (\cref{fig2}) combined with the interface SHA (\cref{fig4}) gives rise to a peak in $j_c^{\perp}(z)$ about $z_{\rm I}$. }
\label{fig5}
\end{figure}

A {\color{blue}fully spin-polarized current} with magnitude $j$ entering an  NM$_1|$NM$_2$ bilayer as sketched in \cref{fig2} undergoes diffusive spin-flipping in each layer as described by \eqref{eq:jsi3} in \cref{subsec:intdis}. If the spin polarization is oriented in the $-x$ direction perpendicular to the current direction $z$ then the spatially decaying spin current $j_s(z) \equiv j_{sx}^z(z)/j$ induces a transverse charge current $j_c^y(z)$ in the $y$ direction, sketched in \cref{fig5}. For two layers labelled $i=1,2$, the normalized charge current is given by $j_c^y(z) = \Theta_i \, j_{si}(z)$. At the interface, the abrupt decay in spin current that is called spin memory loss, combined with an interface SHA $\Theta_{\rm I}$ yields a peak $j_c^y(z_{\rm I}) = \Theta_{\rm I} \bar{J}_s^{\rm I} \delta(z-z_{\rm I})$ where $\bar{J}_s^{\rm I}$ is the effective spin current at the interface. Following Wang {\em et al.} \cite{WangL:prl16}, we integrate the spin current and the transverse charge current in ${\rm NM_2}$ from the interface at $z=z_{\rm I}=0$ out to a distance $L$. As a function of $L$, the total spin current is
\begin{subequations}
\begin{eqnarray}
\!\!\!\!\!\!\!\! \bar{J}_s(L)= \! \int_0^L \!\! j_s(z) dz 
&\approx& \! \int_0^L \!\!\! \big( \bar{J}_s^{\rm I} \delta(z) + j_s^{0+} e^{-z/l_2} \big) dz   \label{eq:10a} \\
&=&\bar{J}_s^{\rm I}+ j_s^{0+} l_2(1-e^{-L/l_2} ),
\label{eq:10b}
\end{eqnarray}
\label{eq:10}
\end{subequations}
where $j_s^{0+}$ is defined as the value of the exponentially decaying current extrapolated to the interface at $z=z_{\rm I}$ from the right, see \cref{fig2}, so that 
\begin{equation}
\bar{J}_s^{\rm I} = \int_0^L \!\! j_s(z) dz - j_s^{0+} l_2(1-e^{-L/l_2}).
\end{equation}
The total transverse charge current induced by the spin current is
\begin{equation}
\bar{J}_c(L)=\Theta_{\rm I}\bar{J}_s^{\rm I} + \Theta_2 \, j_s^{0+} l_2(1-e^{-L/l_2} )
\label{eq:13}
\end{equation}
and an effective SHA due to the ISHE can be defined as
\begin{equation}
\Theta_{\rm eff}(L) \equiv \frac{\bar{J}_c(L)}{\bar{J}_s(L)}
    =\frac{\Theta_{\rm I}\bar{J}_s^{\rm I} + \Theta_2 j_s^{0+} l_2(1-e^{-L/l_2} )}
          {\bar{J}_s^{\rm I} + j_s^{0+} l_2(1-e^{-L/l_2} )}.
\label{eq:thetaeff}
\end{equation}
With the exception of $\Theta_{\rm I}$, all quantities on the right hand side of this expression can be determined independently: $\Theta_2$ and $l_2$ from calculations for bulk NM$_2$, $\bar{J}_s^{\rm I}$ and $j_s^{0+}$ by fitting $j_s(z)$ to the form \eqref{eq:10b}. To extract $\Theta_{\rm I}$, we evaluate $\Theta_{\rm eff}$ from the numerically calculated spin and transverse charge currents and then vary it to optimize the fit to expression \eqref{eq:thetaeff}. 

The procedure for extracting $\Theta_{\rm I}$ proposed by Wang {\em et al.} \cite{WangL:prl16} only takes the contribution from the right side of the interface at $z_{\rm I}=0 $ into account. To account for the interface contribution from both sides, a generalized procedure is derived in \Cref{sec:app}.

\section{Calculations}
\label{sec:Calculations3}

Within the framework of density functional theory \cite{Hohenberg:pr64, Kohn:pr65}, we solve the quantum mechanical scattering problem \cite{Datta:95} for a general two terminal $\mathcal{L}|\mathcal{S}|\mathcal{R}$ configuration consisting of an NM$|$NM$'$ scattering region ($\mathcal{S}$) embedded between ballistic left ($\mathcal{L}$) and right ($\mathcal{R}$) leads (Au or Pt) using a wave-function matching method \cite{Ando:prb91} implemented \cite{Xia:prb06, Zwierzycki:pssb08} with a tight-binding (TB) muffin-tin orbital (MTO) basis in the atomic spheres approximation (ASA) \cite{Andersen:prl84, Andersen:85, Andersen:prb86} and generalized to include SOC and noncollinearity \cite{Starikov:prl10, Starikov:prb18} as well as temperature induced lattice and spin disorder \cite{LiuY:prb11, LiuY:prb15}. The solution yields the scattering matrix $S$, from which we can directly calculate the conductance \cite{Landauer:IBM57, Datta:95}, as well as the full quantum mechanical wave function throughout the scattering region with which we can calculate position dependent charge and spin currents \cite{WangL:prl16, Wesselink:prb19, Nair:prb21a}. Atomic sphere (AS) potentials for Au, Pt and Pd are generated using the Stuttgart TB-LMTO code. Scattering calculations are carried out with an $spd$ orbital basis and two center terms in the SOC Hamiltonian \cite{Starikov:prb18, Wesselink:prb19} with tests carried out with three center terms. In all the calculations that follow, the transport direction is along $z$ and the atomic layers correspond to fcc [111] planes.

\subsection{Lattice mismatch: supercells}

\begin{figure}
\includegraphics[width=1.0\linewidth]{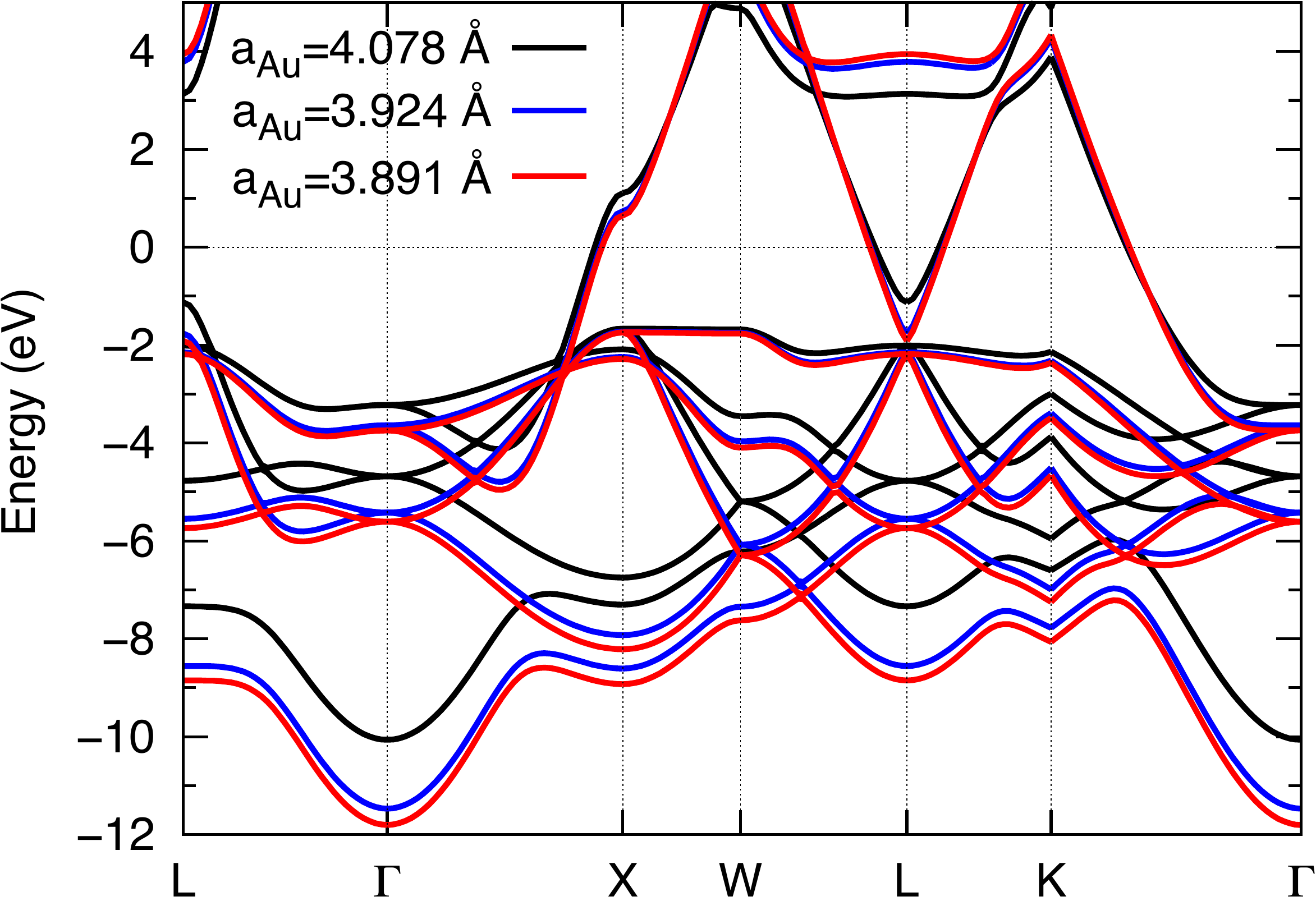}
\caption{Band structure of uncompressed Au (black) and for Au compressed to have the lattice constant of Pt (blue) and Pd (red).}
\label{fig6}
\end{figure}

The lattice constant of Au is 4.078~{\AA}, that of Pt 3.924~{\AA} and of Pd 3.891~{\AA} \cite{Ibach:95}. To construct Au$|$Pt and Au$|$Pd bilayers, the unit cell areas of the two materials should be equal. This can be achieved by using lateral supercells \cite{LiuY:prl14, WangL:prl16, Starikov:prb18}. However, because of its simple nearly-free-electron like nature, the electronic structure of Au does not change qualitatively close to the Fermi energy (shown in \cref{fig6}) when Au is compressed to make it match the lattice constants of Pd and Pt. This makes it much easier to study how modelling disorder in a lateral supercell depends on the supercell size, so we begin by adopting this simpler procedure. The effect of this approximation on the interface transport parameters will be explicitly examined with fully relaxed lateral supercells in \cref{sssubsec:lm} and \cref{sssubsec:lm2} and the results collected for easy comparison in \cref{tab:PtPd}. In \cite{Wesselink:prb19}, we studied the dependence of the spin-flip diffusion length and SHA in Pt on the lateral supercell size used to model thermal disorder. We concluded that a 7$\times$7 supercell was sufficient to obtain adequately converged spin currents and derived parameters. For the bilayer calculations presented in this chapter, we have tested how spin currents passing through a Au$|$Pt interface depend on the supercell size and found that 7$\times$7 is sufficient to describe interface parameters as well. Thus the calculations presented in the Results section are carried out with 7$\times$7 supercells unless stated otherwise. The Brillouin zone (BZ) sampling used for this supercell is 32$\times$32 $k$ points corresponding to an equivalent sampling of 224$\times$224 $k$ points for a 1$\times$1 unit cell.

\subsection{Thermal disorder}
\label{subsec:thermal disorder}

A frozen thermal disorder scheme \cite{LiuY:prb11, LiuY:prb15} is used to model the NM$|$NM$'$ bilayer systems at finite temperatures in the range 200-400 K. We use an uncorrelated Gaussian distribution for the displacements of atoms from their equilibrium lattice positions that is characterized in terms of a root mean square displacement $\Delta$. For each material (NM = Au, Pd, Pt) and temperature $T$, $\Delta_{\rm NM}(T)$ is chosen so as to reproduce the experimental resistivity at that temperature \cite{HCP90}. On constructing the NM$|$NM$'$ bilayer, $\Delta_{\rm NM}(T)$ and $\Delta_{\rm NM'}(T)$ are used for each material to generate multiple configurations with the required thermal disorder at temperature $T$. All the data that will be presented result from averaging over 20 such configurations.

\section{Results}
\label{sec:Results3}

We illustrate the methods described in the previous two sections with results calculated  for the Au$|$Pt interface at 300 K: for $AR_{\rm I}$ in \cref{subsec:intRAuPt}, for $\delta$ in \cref{subsec:delta}, for the interface SHE in \cref{subsec:sha} and for $\Theta_{\rm I}$ from the ISHE in \cref{subsec:thetaI}. The Au$|$Pt interface parameters are compared to those calculated for Au$|$Pd at 300 K in \cref{subsec:AuPt_AuPd}. Finally, the dependence of $AR_{\rm I},~\delta$ and  $\Theta_{\rm I}$ on temperature is presented in \cref{subsec:temp}.

\subsection{Au$|$Pt: interface resistance}
\label{subsec:intRAuPt}

\begin{figure}[t]
\includegraphics[width=8.6cm]{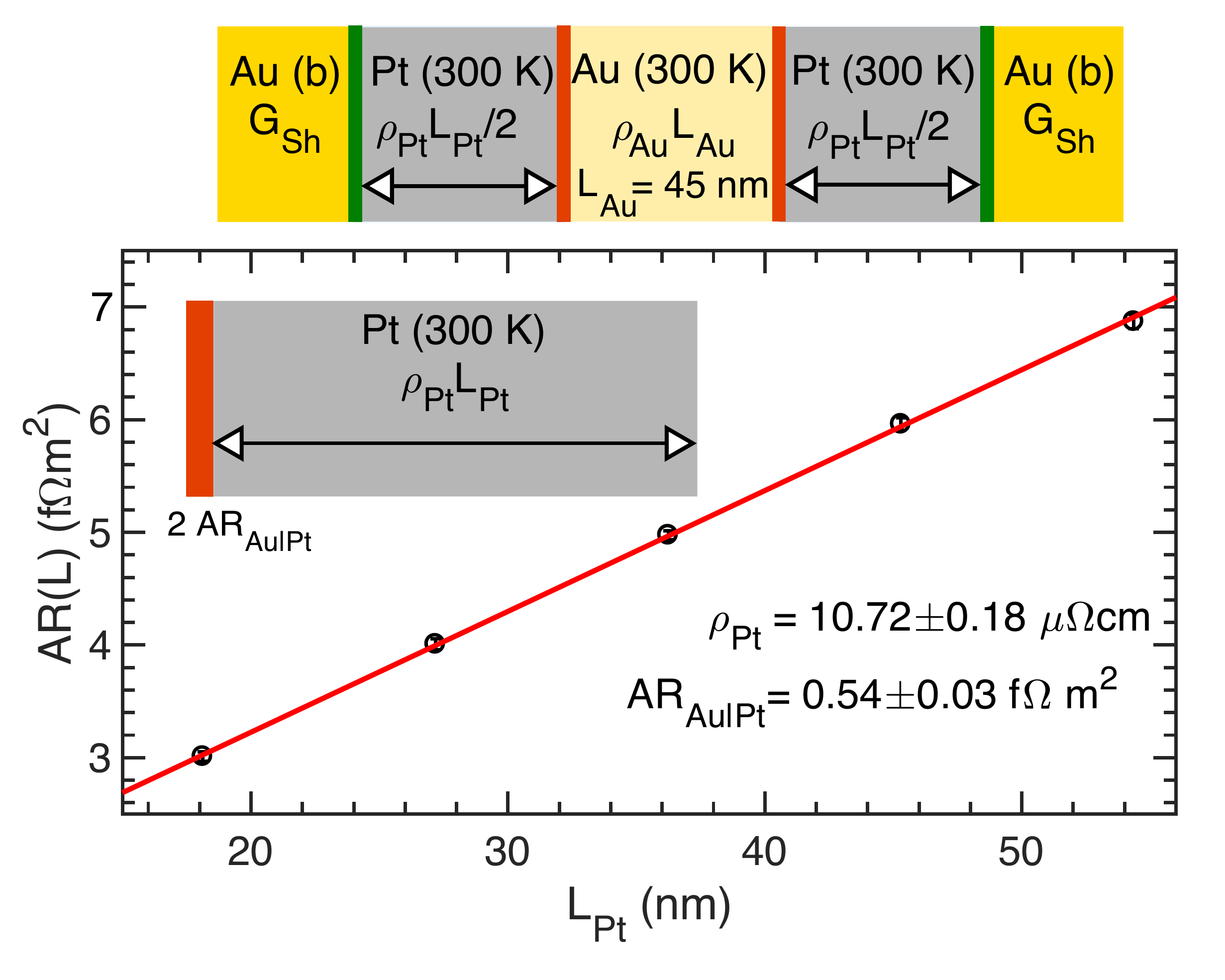}
\caption{Total resistance of a diffusive Pt$|$Au$|$Pt trilayer, Pt$(L_{\rm Pt}/2)|$Au$(L_{\rm Au})|$Pt$(L_{\rm Pt}/2)$ sandwiched between ballistic Au leads as a function of the total Pt thickness $L_{\rm Pt}$ for a fixed Au thickness $L_{\rm Au}$ = 45 nm. To extract $AR_{\rm  Au|Pt}$ (red), contributions from diffusive Au (pale yellow), ballistic Au ($G_{\rm Sh}$, yellow), and the interface resistance between ballistic Au and Pt (green) are calculated separately (see \cref{fig8}) and subtracted. $\rho_{\rm Au}$ is also determined separately. The intercept of $AR(L)$ at $L=0$ yields $AR_{\rm Au|Pt}$.}
\label{fig7}
\end{figure}

The interface resistance $AR_{\rm Au|Pt}$ is extracted in a two step procedure. We first calculate the total resistance for a symmetric Pt$|$Au$|$Pt trilayer embedded between ballistic Au leads for a variable length $L_{\rm Pt}$ of Pt and fixed length of Au, $L_{\rm Au}=45$~nm. Both $L_{\rm Pt}$ and $L_{\rm Au}$ should be much longer than the respective mean free paths so that the total areal resistance for the scattering region can be expressed in terms of the series resistor model, sketched at the top of \cref{fig7}, as
\begin{eqnarray}
AR(L_{\rm Pt},L_{\rm Au})&=\rho_{\rm Pt}L_{\rm Pt}+\rho_{\rm Au}L_{\rm Au}+2AR_{\rm Au|Pt}\nonumber\\
&+2AR_{\rm Pt|Au(b)}+1/G_{\rm Sh}.
\label{eq:LNFNL3}
\end{eqnarray}
Here, $AR_{\rm Au|Pt} \equiv AR_{\rm I}$ is the interface resistance we are interested in, $AR_{\rm Pt|Au(b)}$ is the interface resistance between Pt and the ballistic Au lead, and $G_{\rm Sh}$ is the Sharvin conductance of the Au lead. In separate calculations for a variable thickness $L_{\rm Pt}$ of Pt embedded between Au leads, shown in \cref{fig8}, the total areal resistance  
\begin{equation}
AR(L_{\rm Pt})= \rho_{\rm Pt}L_{\rm Pt} + 2AR_{\rm Pt|Au(b)}+1/G_{\rm Sh}
\label{eq:LNL3}
\end{equation}
is determined. Fitting $AR(L_{\rm Pt})$ to \eqref{eq:LNL3} yields $\rho_{\rm Pt}$ as the slope and the final two terms as the intercept. A similar calculation for diffusive Au yields $\rho_{\rm Au}$. 
We subtract the contributions $2AR_{\rm Pt|Au(b)}+1/G_{\rm Sh}$ as well as $\rho_{\rm Au}L_{\rm Au}$ from $AR(L_{\rm Pt},L_{\rm Au})$ and plot the remainder, $\rho_{\rm Pt}L_{\rm Pt} + 2AR_{\rm Au|Pt}$, in \cref{fig7}. Linear fitting yields the $L_{\rm Pt}=0$ intercept  $AR_{\rm Au|Pt} = 0.54 \pm 0.03 \,{\rm f}\Omega\,{\rm m}^2$ at 300 K.

\subsection{Au$|$Pt: ${\boldmath \delta}$}
\label{subsec:delta}

To calculate $\delta$ for a Au$|$Pt interface, we inject a {\color{blue}fully spin-polarized current} from a ballistic Au$\uparrow$ lead into a Au$|$Pt bilayer sandwiched between Au leads. The diffusive Au slab into which the spins are injected should be thick enough to avoid artifacts arising from ballistic transport but sufficiently thin that a substantial spin current still enters Pt after spin flipping has occurred at the interface. We carried out tests with various lengths of Au (50, 100, 150 atomic layers) keeping Pt fixed at 150 layers and found that the final results for $\delta$ were not affected by this choice. The results presented here are for 50 layers ($\sim10$ nm) of Au and 150 layers ($\sim30$ nm) of Pt that we denote Au(10)$|$Pt(30). Both Au and Pt are modelled at 300 K using the rms displacements discussed in \cref{subsec:thermal disorder}. 

The left lead is made to be ``half-metallic'' by lifting the bands of one spin channel above the Fermi energy so that a {\color{blue}fully spin-polarized current} flows into the bilayer. As seen in \cref{fig9}, the spin current decays rather slowly in Au reflecting the large value we found for $l_{\rm Au}$. At the Au$|$Pt interface, we see a sharp decrease in the spin-current as it enters Pt which is a clear indication of spin-memory loss. The spin current then decays exponentially towards zero in Pt. Giving values of the spin current close to the interfaces less weight, we fit $j_s(z)$ piecewise in Au and Pt using \eqref{eq:js3} and extrapolate the fitted curves to the interface at $z_{\rm I}$ to obtain $j_{s,{\rm Au}}(z_{\rm I})$ and $j_{s,{\rm Pt}}(z_{\rm I})$. In \cref{fig9}, we shift the origin of the $z$ axis so that $z_{\rm I}=0$ for convenience. This does not impact the boundary conditions considered in \cref{subsec:intdis} as the absolute value of $z_{\rm I}$ does not enter \eqref{eq:delta}.

\begin{figure}[t]
\includegraphics[width=8.6cm]{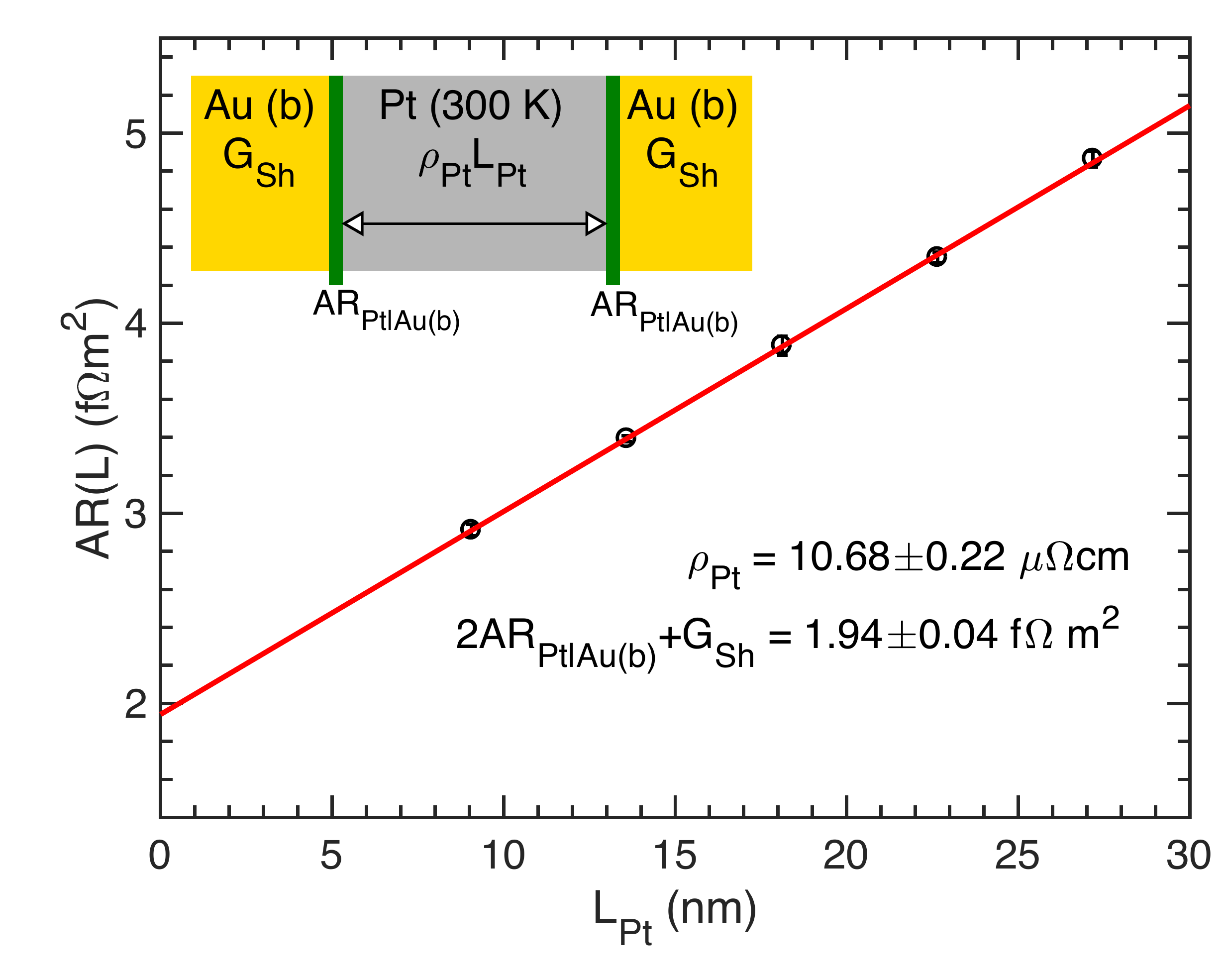}
\caption{Total resistance for diffusive Pt sandwiched between ballistic(b) Au leads as a function of the Pt thickness $L_{\rm Pt}$. A linear fit $AR(L)$ yields $\rho_{\rm Pt}$ as the slope; the intercept is a sum of interface and Sharvin contributions.}
\label{fig8}
\end{figure}

\begin{figure}[t]
\includegraphics[width=8.6cm]{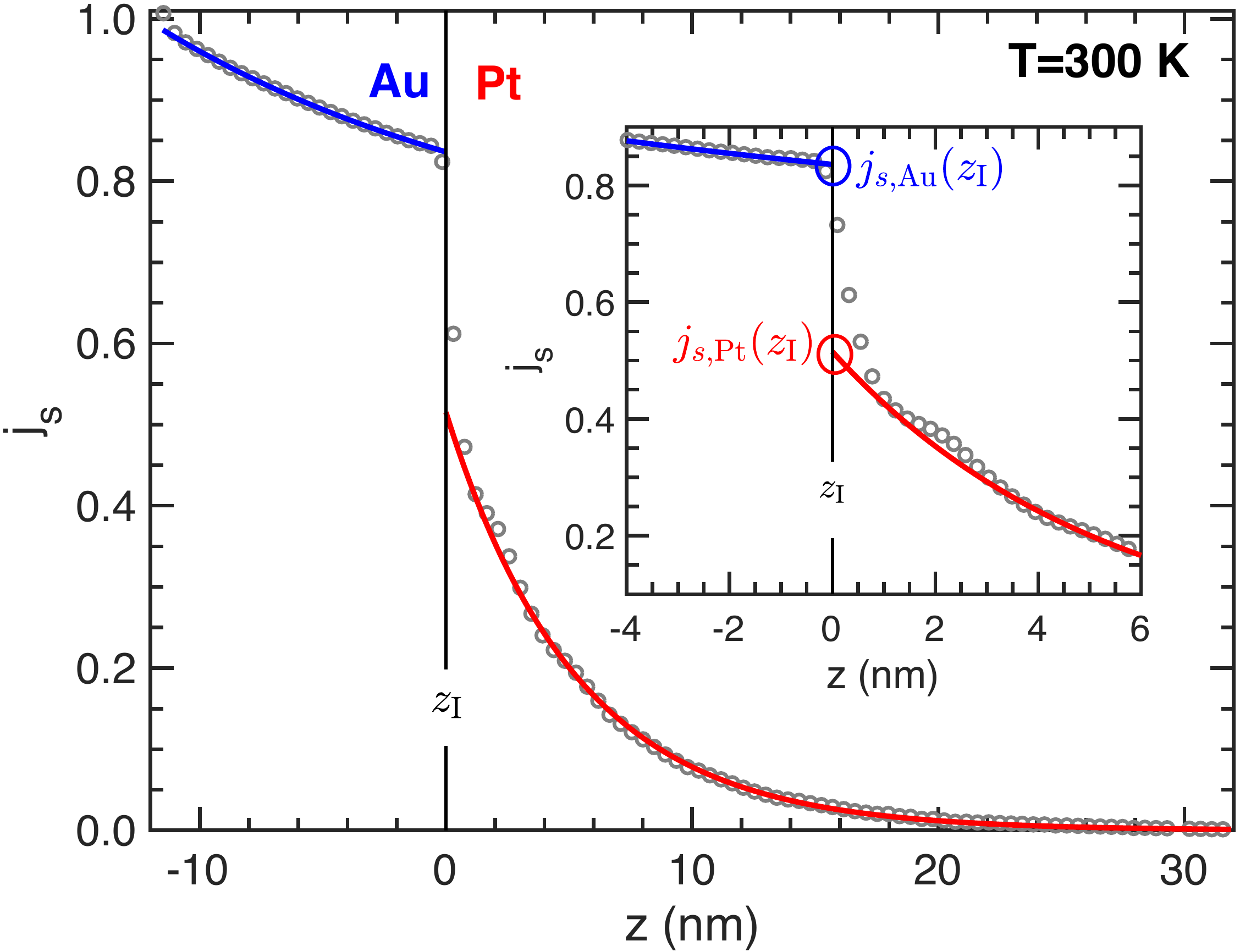}
\caption{A {\color{blue}fully spin-polarized current} $j_s$ injected at 300 K from the left lead into a Au(10)$|$Pt(30) bilayer sandwiched between Au leads decays exponentially in Au and in Pt. The numbers in brackets are lengths in nm. The solid lines indicate fits for $j_s$ in individual layers using \eqref{eq:jsi3}. The change in the spin current in the vicinity of the interface that is not ``bulk-like" indicates the interfacial SML $\delta$. Inset: Exploded view of the interface. $j_{s,{\rm Au}}(z_{\rm I})$ and $j_{s,{\rm Pt}}(z_{\rm I})$ indicate the values of the bulk spin currents extrapolated to $z_{\rm I}$ from the Au side and Pt side, respectively. $\delta$ is extracted using these values and \eqref{eq:delta}. 
}
\label{fig9}
\end{figure}

We rewrite \eqref{eq:delta} for the Au$|$Pt interface as
\begin{equation}
	\frac{j_{s,\rm Au}(z_{\rm I})}{j_{s,\rm Pt}(z_{\rm I})}=\cosh{\delta}+\delta \sinh{\delta}~\frac{\rho_{\rm Pt} l_{\rm Pt}}{AR_{\rm Au|Pt}}.
\label{eq:deltaAuPt}	
\end{equation}
In our previous work \cite{Wesselink:prb19}, we extracted a value of $l_{\rm Pt}=5.25\pm0.05$ nm at T=300~K using the same lateral supercell, basis and two center terms. As shown in \cref{fig8}, $\rho_{\rm Pt}=10.68\pm0.22~\mu\Omega\rm cm$. The only unknown in \eqref{eq:deltaAuPt} is $\delta$. Substituting all the other parameters in \eqref{eq:deltaAuPt}, $\delta$ can be extracted using a numerical root finder. For the Au$|$Pt interface at 300 K, we find $\delta=0.62\pm0.03$. The error bar is evaluated by taking into account the spread of all the input parameters as described by their respective error bars. 

The clean, lattice-matched interface is ideal for performing systematic studies to investigate the effect of changing the temperature on $AR_{\rm I}$ and $\delta$; this will be done in \cref{subsec:temp}. Before doing so, we should remember that real interfaces are not sharp and we need to consider the effects of intermixing as well as lattice mismatch between Au and Pt. This we will do in the following paragraphs where the values of $AR_{\rm Au|Pt} = 0.54 \pm 0.03 \,{\rm f}\Omega\,{\rm m}^2$ and $\delta=0.62\pm0.03$ obtained for an ideal lattice matched interface at room temperature will serve as reference values in \cref{tab:PtPd}.

\subsubsection{Interface mixing}

To study the effect of interface mixing, we consider $N$ atomic layers at the interface to consist of a Au$_{50}$Pt$_{50}$ random alloy for which the Au and Pt AS potentials are calculated using the coherent potential approximation (CPA) \cite{Soven:pr67, Turek:97}. These AS potentials are distributed randomly in the $N$ interface layers so as to maintain the correct stoichiometry. The thermal disorder is modelled using the average of the room temperature values of $\Delta_{\rm Au}$ and $\Delta_{\rm Pt}$. 

\begin{figure}[t]
\includegraphics[width=8.6cm]{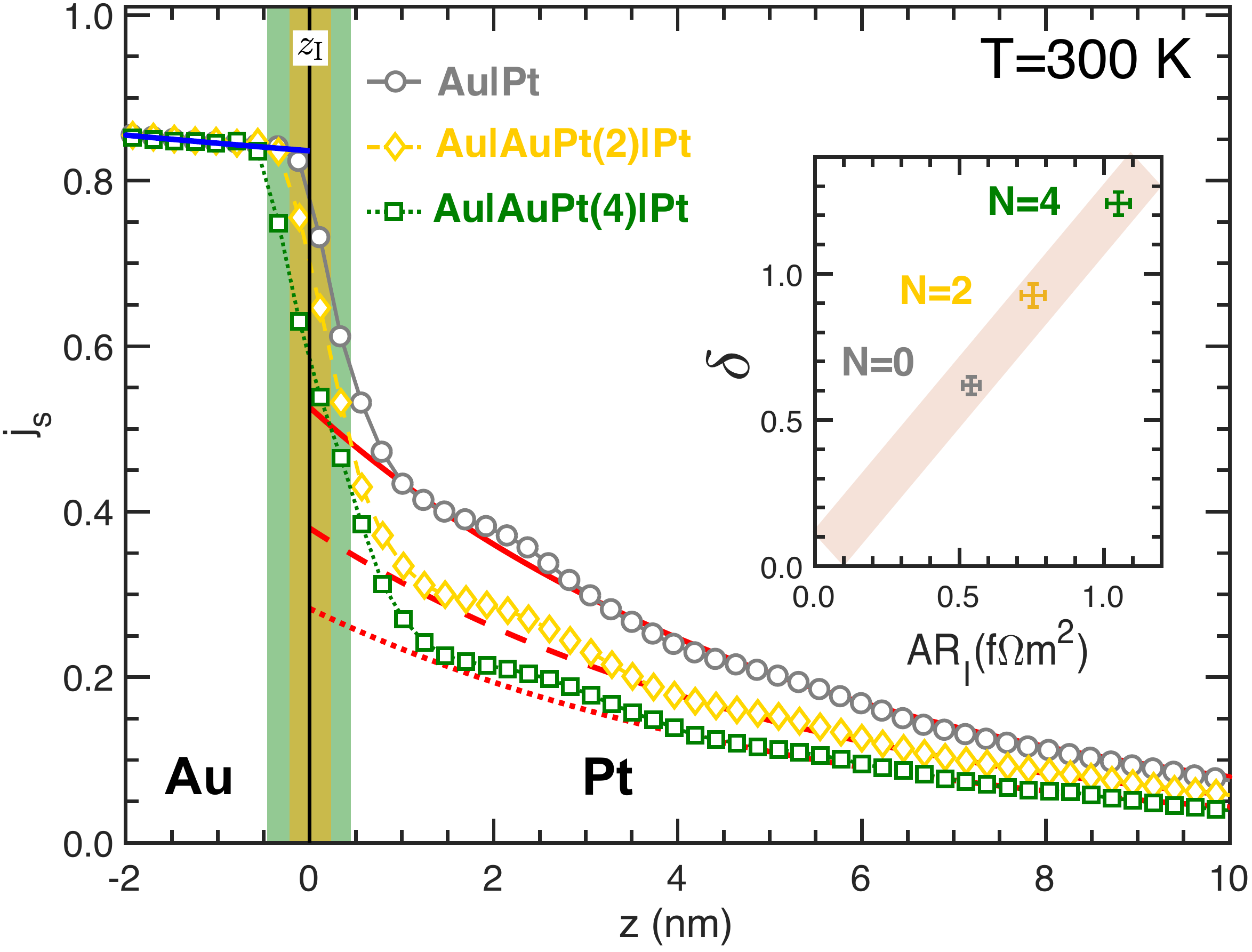}
\caption{A {\color{blue}fully spin-polarized current} $j_s(z)$ is injected into a Au$|$Pt bilayer  with: a sharp interface (vertical black line), 2 layers of $\rm Au_{50}Pt_{50}$ interface (yellow shaded region) and 4 layers of $\rm Au_{50}Pt_{50}$ interface (green shaded region) between them. The calculated spin currents $j_s(z)$ for the three cases are shown as gray circles, yellow diamonds and green squares respectively. The solid blue line indicates a fit to the VF equation in Au. The solid, dashed and dotted red lines indicate fits to the VF equation in Pt for Au$|$Pt, Au$|\rm Au_{50}Pt_{50}(2)|$Pt and Au$|\rm Au_{50}Pt_{50}(4)|$Pt respectively. Inset: $\delta$ vs $AR_{\rm I}$ for $N$=0,2,4 interface layers of mixed $\rm Au_{50}Pt_{50}$.}
\label{fig10}
\end{figure}

The results we obtain for the spin current $j_s(z)$ for this model of interface disorder and the corresponding values of $\delta$ and $AR_{\rm I}$ at 300 K are shown in \cref{fig10} for $N=0,2,4$. The spin current incident on the interface only differs from that in the ideal, sharp interface case in the intermixed layers themselves (yellow for $N=2$, green for $N=4$) where $j_s(z)$ decreases more rapidly with increasing disorder corresponding to larger values of $\delta$ and $AR_{\rm I}$ (inset). At a lattice-matched, commensurable and clean Au$|$Pt interface, crystal momentum parallel to the interface is conserved and electron scattering only involves Bloch states with the same $\mathbf k_\|$. Intermixing (and thermal disorder) breaks momentum conservation and allows ${\bf k_\| \rightarrow k'_\|}$ scattering. The higher scattering rate results in a higher spin-flipping probability and hence larger $\delta$ and $AR_{\rm I}$ for the intermixed interfaces. Moreover, conduction electrons at the Fermi level in Au are only weakly affected by SOC. As $d$ states in Pt, they become very susceptible to the large SOC. The interatomic mixing effectively increases the region where conduction electrons experience large SOC and this therefore increases the SML. In the inset we see that $\delta \propto AR_{\rm I}$ with the factor $AR_{\rm I}/\delta$ having a constant value $\sim 0.85 \,{\rm f}\Omega {\rm m}^2$ corresponding to $\rho_{\rm I} l_{\rm I}$ being constant.

\begin{figure*}[t]
\includegraphics[width=17.8cm]{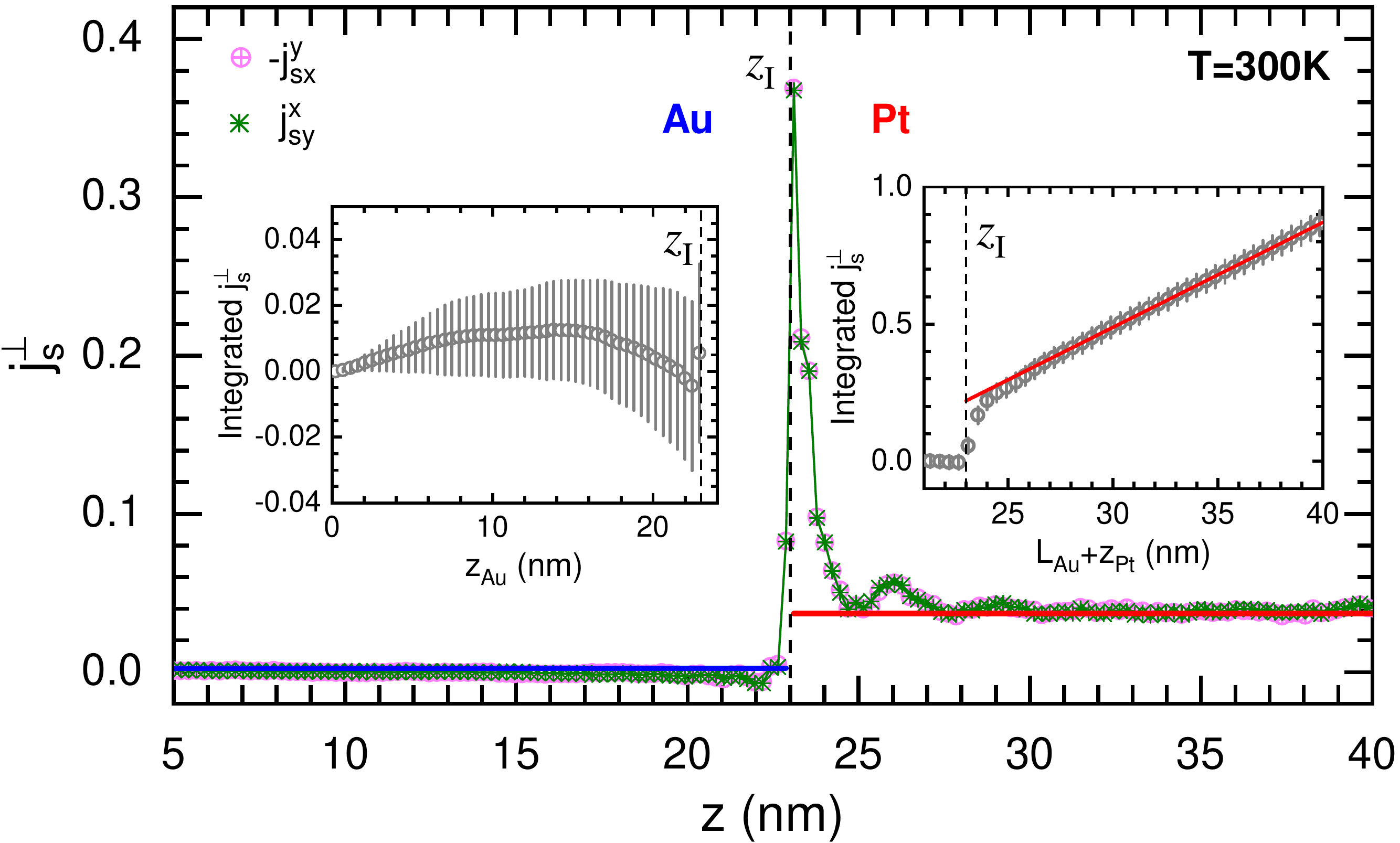}
\caption{SHE in a Au(20~nm)$|$Pt(20~nm) bilayer at 300 K embedded between Au and Pt leads on the left and right, respectively showing transverse spin currents $-j_{sx}^y$ as pink circles and $j_{sy}^x$ as green crosses. The blue and red horizontal lines show the values of the bulk SHAs of Au, $\Theta_{\rm Au}=0.25\%$ and of Pt, $\Theta_{\rm Pt}=3.7\%$ calculated separately for bulk diffusive Au and Pt. Left inset: Integrated transverse spin current in Au. Right inset: Integrated transverse spin current in Pt. The  solid red line indicates the fit obtained from the integrated spin current in Pt. The intercept at $z_{\rm I}$ shows the contribution from the interface. 
}
\label{fig11}
\end{figure*}

\subsubsection{Lattice mismatch} 
\label{sssubsec:lm}
To study the effect lattice mismatch has on the interface parameters, we calculate them for a (111) Au$|$Pt interface where both Au and Pt have their equilibrium bulk volumes given by $a_{\rm Au}=4.078\,$\AA\ and $a_{\rm Pt}=3.924\,${\AA}. A (111) oriented $5\times5$ unit cell of Au matches with a similarly oriented $3\sqrt{3}\times3\sqrt{3}$ unit cell of Pt to better than 0.02\%. The unit cells need to be rotated with respect to each other to make them coincide. For this fully relaxed Au$|$Pt geometry, we repeat our calculations at 300~K and find $AR_{\rm I}=0.81 \pm 0.04\,{\rm f}\Omega\,{\rm m}^2$ and $\delta=0.81\pm0.05$, \cref{tab:PtPd}. Both interface parameters obtained with the Au lattice in equilibrium are larger than those obtained with compressed Au; the interface is more pronounced. The same trend will be found for the corresponding Au$|$Pd interfaces. We attribute this to the lack of conservation of transverse momentum and greater ${\bf k_{\|} \rightarrow k'_{\|}}$ scattering in the absence of lattice matching.

\subsection{Au$|$Pt: interface SHE}
\label{subsec:sha}

In \cref{subsec:shaint} we discussed the qualitative behaviour of the spin Hall effect in the bulk of two nonmagnetic materials and how it abruptly changes at an interface, \cref{fig5}. We now pass an unpolarized charge current through a 300~K diffusive Au(20 nm)$|$Pt(20 nm) bilayer sandwiched between a ballistic Au lead on the left and Pt lead on the right, chosen to minimize interface contributions from the leads. Transverse spin currents generated by the SHE in the two materials and at their interface are shown as pink circles ($j_{sy}^x$) and green crosses ($-j_{sx}^y$) in \cref{fig11}; for the axially symmetric CPP geometry, $j_{sy}^x=-j_{sx}^y$. The horizontal blue and red lines show the values of the bulk SHAs of Au, $\Theta_{\rm Au}=0.25\%$, and of Pt, $\Theta_{\rm Pt}=3.7 \%$, determined separately for homogeneous scattering regions. Sufficiently far from the interface, spin currents are seen to obey their bulk behaviour in both materials. For Pt, this happens very quickly; for Au it takes much longer suggesting the spin-flip diffusion length as the relevant length scale. In a region of $\sim 5\,$nm about the interface at $z_{\rm I}$, we see a clear deviation from bulk behaviour in both Au and Pt that culminates in a large interface spin Hall contribution \footnote{We attribute the oscillations in $j_c^{\perp}(z)$ to Fermi surface nesting effects, quantum effects that are absent in semiclassical transport formalisms. }. To quantitatively describe this sharp peak, we integrate the transverse spin currents in the bilayer starting from the interface with the left Au lead at $z=0$ up to the Au$|$Pt interface at $z=z_{\rm I}$ ($z_{\rm Au}=L_{\rm Au}$ in the left-hand inset) and onward to the interface with the right Pt lead at $z=L_{\rm Au}+L_{\rm Pt}$ (right-hand inset). 

\begin{figure}[t]
\includegraphics[width=8.6cm]{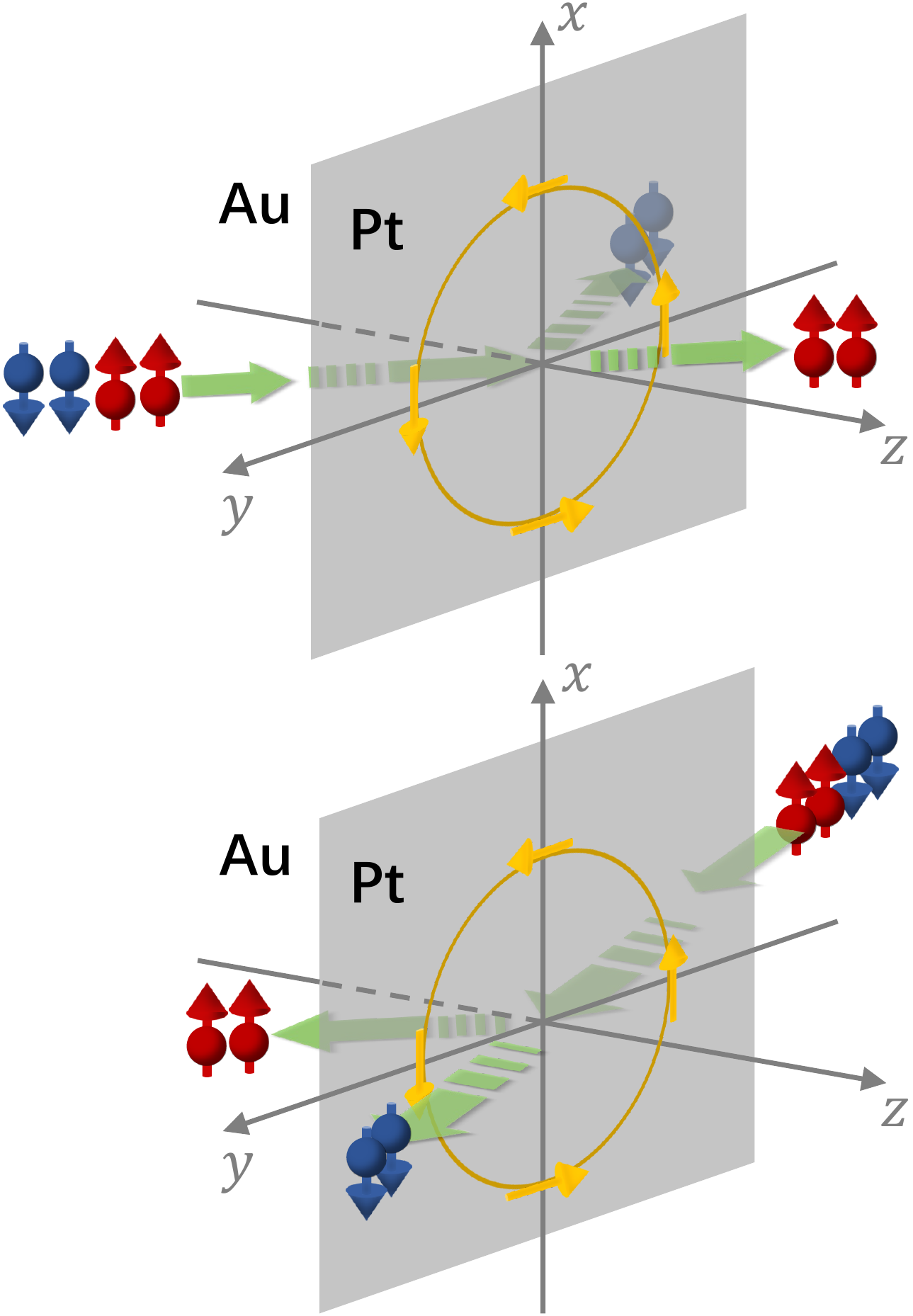}
\caption{Illustration of the reflection of conduction electrons at the Au$|$Pt interface resulting from a Rashba-type spin-orbit interaction. The yellow circle represents a surface of constant energy in a ($k_x, k_y$) plane appropriate to a (111) surface of Au \cite{LaShell:prl96}. The arrows on the circle indicate the effective Rashba field at the Fermi level which depends on the local momentum. The spin- and momentum-dependent reflection results in a transverse spin current polarized along $x$ and flowing along $+y$ on the Au side of the interface, i.e. $j_{sx}^y$, which is opposite to the spin Hall current with a positive SHA. 
} 
\label{fig11Rashba}
\end{figure}

The integral $\int_0^{z_{\rm Au}} dz' (j_{sy}^x-j_{sx}^y)$ plotted in the left-hand inset of \cref{fig11} shows the integrated spin current increasing from zero up to a certain value of $z_{\rm Au}$ before decreasing again to essentially zero close to the interface at $z_{\rm I}$. If we add the total integrated contribution from Au and continue to integrate through Pt, the result is $\int_0^{L_{\rm Au}} dz' (j_{sy}^x-j_{sx}^y) + \int_{z_{\rm I}\equiv L_{\rm Au}}^{L_{\rm Au}+z_{\rm Pt}} dz' (j_{sy}^x-j_{sx}^y)$ and it is shown in the right-hand inset. It can be fitted with a straight line whose slope is just the value we calculate independently for bulk Pt, $\Theta_{\rm Pt}=3.7\pm0.1\%$. The finite intercept 0.22 yields the contribution from the interface spin Hall effect in units of nm. To extract a dimensionless interface SHA, $\Theta_{\rm I}$, we make use of the charge currents generated by the ISHE in Au$|$Pt in the next section. 


\begin{figure}[t]
\includegraphics[width=8.6cm]{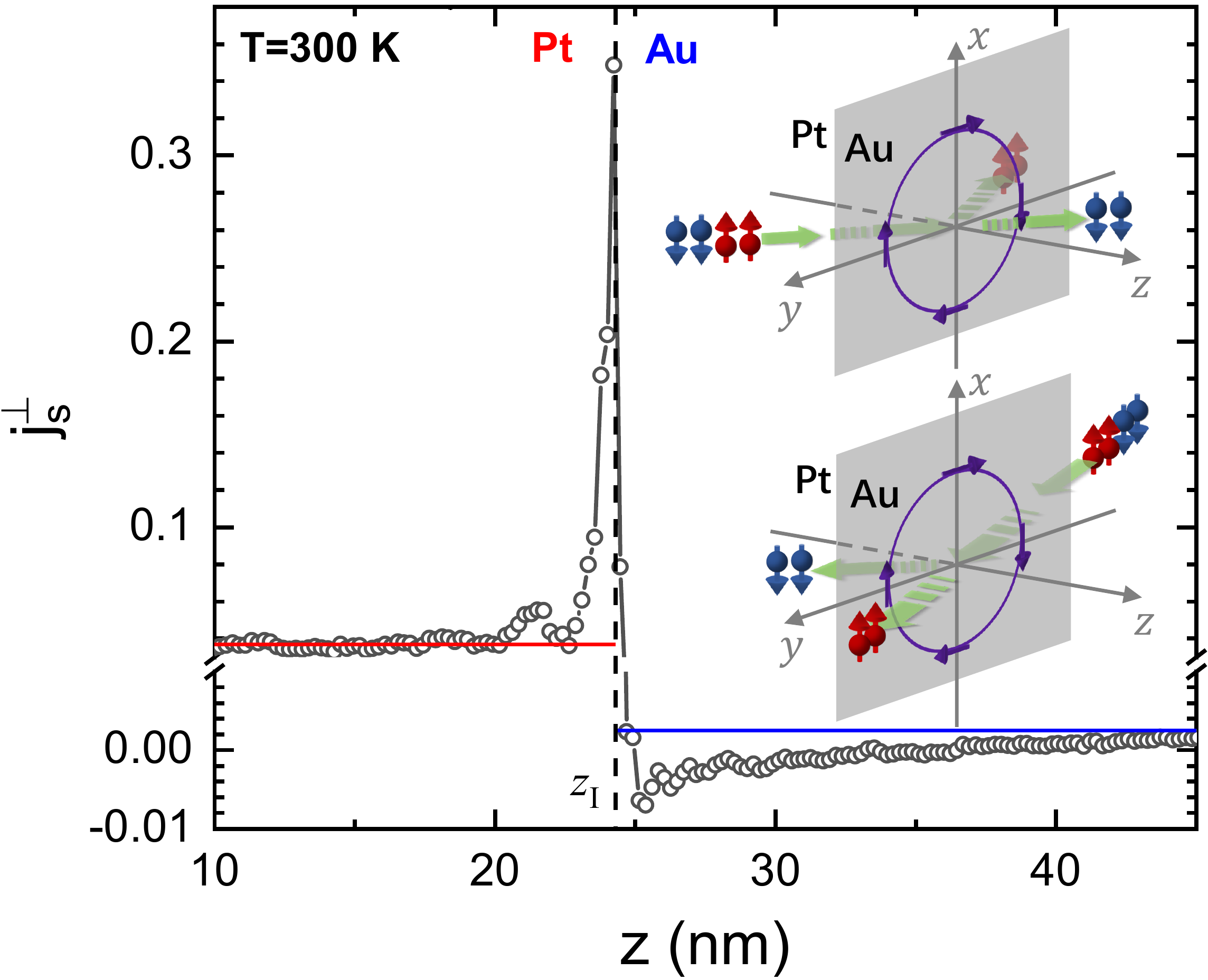}
\caption{Calculated transverse spin current for a charge current $j_c$ injected from Pt to Au. The blue and red horizontal lines show the values of the bulk SHAs of Au, $\Theta_{\rm Au}=0.25\%$ and of Pt, $\Theta_{\rm Pt}=3.7\%$ calculated separately for bulk diffusive Au and Pt. Note the magnified axis for the negative range. Inset: sketch to illustrate the transverse spin current induced by the filtering effect.} 
\label{fig11PtAu}
\end{figure}

\begin{figure*}[t]
\includegraphics[width=17.8cm]{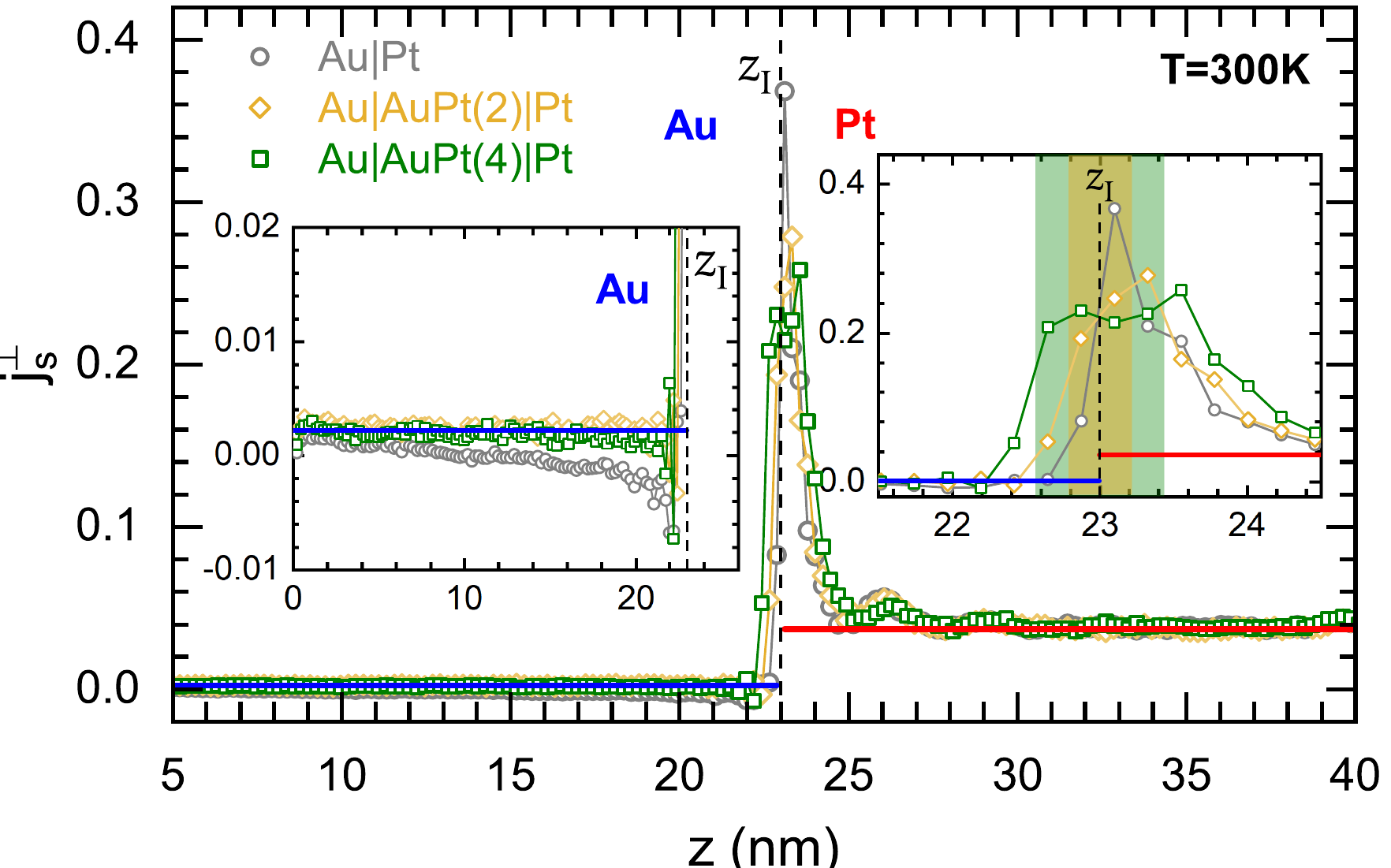}
\caption{SHE at 300 K in a Au(20~nm)$|$Pt(20~nm) bilayer with: a sharp interface (vertical black dashed line), 2 layers of $\rm Au_{50}Pt_{50}$ interface alloy (yellow shaded region in right inset) and 4 layers of $\rm Au_{50}Pt_{50}$ interface alloy (green shaded region in right inset) embedded between ballistic Au (left) and Pt (right) leads. The resulting transverse spin currents $j^{\perp}_s=(j^x_{sy}-j^y_{sx})/2$ for the three cases are shown as gray circles, yellow diamonds and green squares respectively. The blue and red horizontal lines show the values of the bulk SHAs of Au, $\Theta_{\rm Au}=0.25\%$ and of Pt, $\Theta_{\rm Pt}=3.7 \%$ calculated separately for bulk diffusive Au and Pt. An exploded view about the interface is shown in the right inset. In the left inset the vertical scale focuses attention on the Au side of the interface.
}
\label{fig11disorder}
\end{figure*}

A closer inspection of the transverse spin current at the Au$|$Pt interface suggests that the spin Hall current near the interface on the Au side is slightly negative, i.e., opposite in sign to the bulk SHA of Au. This negative contribution from the Au side is also seen in the left inset of \cref{fig11} where the integrated spin current decreases as the interface at $z_{\rm Au}=L_{\rm Au}$ is approached \footnote{Similar calculations for Cu$|$Pt reproduce the distribution of transverse spin current including the strong and positive interface enhancement on the Pt side and a small negative net SHA contribution on the Cu side that is however larger in magnitude than for Au$|$Pt because bulk Cu has a smaller SHA.}. It can be understood as a filtering effect of a Rashba-type SOC at the Au$|$Pt interface \cite{YuR:prb20, Manchon:natm15} which induces a spin-momentum locking at the Fermi level so that the effective magnetic field experienced by conduction electrons depends on their momenta, as indicated by the arrows on the circles in \cref{fig11Rashba}. 
For a current of electrons flowing in the $z$ direction driven by an external voltage, we consider electrons arriving at the interface (plane) from the Au side with in-plane velocity components along $+y$ (bottom panel) and $-y$ (top panel) with spin up or down with respect to the $x$ axis (red and blue arrows, respectively). 

For electrons with an in-plane velocity component along $-y$ (i.e. top panel), the spin-up electrons (red) find it easier to pass through the interface because the potential barrier they see at the interface is reduced by the Rashba field while that of the spin-down (blue) electrons is increased and they are reflected more \cite{LiS:prb19}. 
For electrons with an in-plane velocity component along $+y$ (i.e. bottom panel) spin-down electrons (blue) have a higher transmission probability while spin-up electrons are reflected relatively more.
This spin-selective reflection leads to transverse spin currents that are in opposite directions on either side of the interface. On the Pt side, an up-spin current flows in the $-y$ direction i.e., it has a positive SHA reinforcing the intrinsic Pt SHE. On the Au side, this spin current subtracts from the intrinsic positive spin Hall current as found in \cref{fig11}. 

The negative contribution resulting from the interface filtering effect is independent of the stacking order of Au and Pt as confirmed by repeating the calculation but now with a charge current $j_c$ injected from Pt into Au. The calculated transverse spin current $j^{\perp}_s=(j^x_{sy}-j^y_{sx})/2 \sim -j^y_{sx}$ is shown in \cref{fig11PtAu}. With the reversed stacking order of Au and Pt, the effective Rashba field keeps its clockwise rotation as seen from the Au side. Therefore, as illustrated in the inset to \cref{fig11PtAu}, the Rashba-type SOC induces a transverse spin current $j_{sx}^y$ (in the $+y$ direction with polarization $+x$) on the Au side, which is opposite to the positive direction of $j^{\perp}_s \sim -j^y_{sx}$ and thus has a negative magnitude. On the Pt side, the reflection results in a positive $j^{\perp}_s$. The calculated $j^{\perp}_s$ shown in \cref{fig11PtAu} is consistent with the above expectation.

\subsubsection{Interface mixing}
When the interface is no longer atomically sharp because of intermixing of NM and NM$'$ atoms in the interface layers, the increased interface resistance found in \cref{subsec:delta} may largely reduce the backflow spin current discussed above. We examine this expectation  by inserting $N$ atomic layers of $\rm Au_{50}Pt_{50}$ random alloy at the interface of the lattice-matched $\rm Au|Pt$ bilayer. The calculated transverse spin currents $j^{\perp}_s=(j^x_{sy}-j^y_{sx})/2$ for $N=0,2,4$ are shown in \cref{fig11disorder}. The large interface spin Hall current in the clean Au$|$Pt interface case is reduced in magnitude but increased in width by interface disorder as shown in the right inset of \cref{fig11disorder}. The contribution that was negative on the Au side for the clean interface is quenched by interface disorder. This is seen more clearly in the left inset of \cref{fig11disorder}. We attribute this quenching to suppression of the effective Rashba field by interface disorder. The increase in the width of the interface spin Hall enhancement means that the interface term is no longer largely on the Pt side of the interface so a quantitative estimate of the interface SHA  with interface mixing must include the contribution from both sides as will be discussed in \Cref{sec:app}.


\subsection{Au$|$Pt: interface ISHE - ${\bf \Theta_{\rm I}}$}
\label{subsec:thetaI}

\begin{figure}[t]
\includegraphics[width=8.6cm]{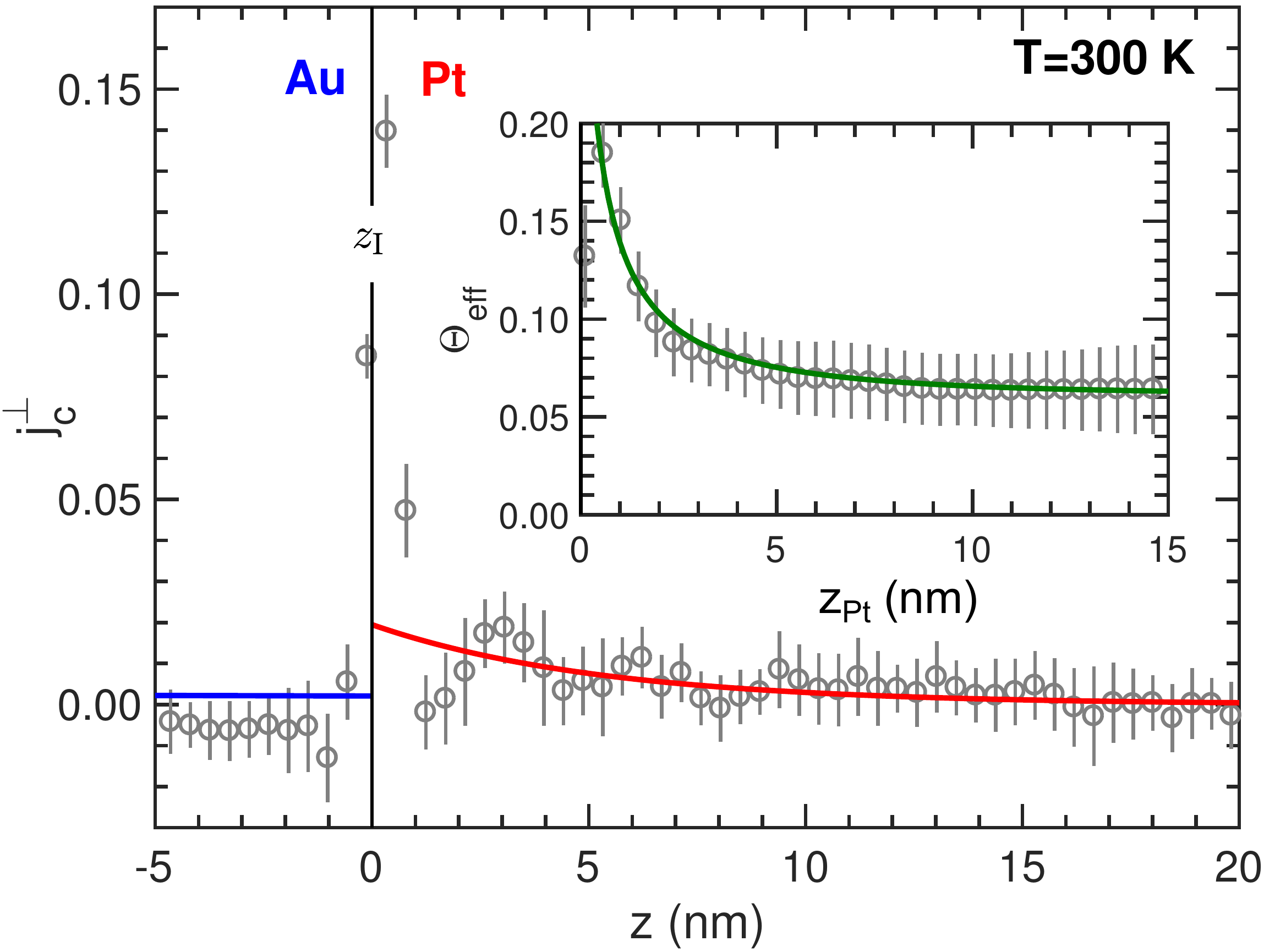}
\caption{$z$ dependence of the ISHE charge current in the $y$ direction (open circles) induced by the spin current polarized in the $-x$ direction shown in \cref{fig9}. The solid blue and red lines indicate the fitted $j_{s,{\rm Au}}$ and $j_{s,{\rm Pt}}$ from \cref{fig9} multiplied by the bulk spin Hall angles: $\Theta_{\rm Au}=0.0025$ and $\Theta_{\rm Pt}=0.037$ respectively. Inset: Effective SHA $\Theta_{\rm eff}$ calculated from the spin and charge currents integrated from $z'_{\rm Pt}=0$ to $z'_{\rm Pt}=z_{\rm Pt}$. The solid green line is the fit obtained using \eqref{eq:thetaeff}.} 
\label{fig12}
\end{figure}

The spin current $j_s(z)$ shown in \cref{fig9} is polarized in the $-x$ direction \footnote{The spin memory loss could depend on this polarization direction. If it does, then the dependence is small and falls within the 5\% accuracy of the present calculations.}. A consequence of the (inverse) spin Hall effect is that such a transversely polarized spin current induces a charge current $j_c^{\perp}(z)$ which is given by the vector product of the current and polarization directions; its magnitude is shown in \cref{fig12}. Far from the interface, this charge current should simply depend on the material-specific SHA $\Theta_i$ as $\Theta_i \, j_{si}(z)$. By comparing this product with the explicitly calculated $j_c^{\perp}(z)$, we can identify departures from the expected bulk behavior and attribute them to the interface. We already fitted $j_{si}(z)$ to \eqref{eq:js3}, resulting in the blue and red solid lines in \cref{fig9}. In \cite{Wesselink:prb19}, we calculated $\Theta_{\rm Pt}=3.7 \pm 0.1 \%$ (using 2-center terms and $spd$ orbitals) at 300 K. Using the same procedure, we find the SHA for Au to be $\Theta_{\rm Au}=0.25 \%$ at 300 K \cite{Nair:prl21}. 

The solid blue and red lines in \cref{fig12} represent $\Theta_{\rm Au} \, j_{s,{\rm Au}}(z)$ and $\Theta_{\rm Pt} \, j_{s,\rm Pt}(z)$, respectively. On the Pt side of the interface, the calculated $j_c^{\perp}(z)$ approaches the expected bulk value $\Theta_{\rm Pt} \, j_{s,\rm Pt}(z)$ as we move away from the interface. Right at the interface, a high and narrow  $j_c^{\perp}(z)$ signals an interface SHA much larger than the Pt bulk SHA. In Au it seems that the transverse charge current $j_c^{\perp}(z)$ has not yet reached its asymptotic bulk value. 
Although the current injected from the Au lead is still almost fully spin-polarized at the Au$|$Pt interface because of the weak spin-flipping in Au, a very small ISHE bulk charge current, shown by the blue line in \cref{fig12}, is expected because of the very small value of $\Theta_{\rm Au} = 0.25\%$. However, the actual $j_c^{\perp}(z)$ is seen to be negative and this can be attributed to the spin- and momentum-dependent reflection by the Rashba-type spin-orbit interaction at the interface by analogy with the negative spin current on the Au side shown in \cref{fig11}.

To extract $\Theta_{\rm I}$, we integrate the calculated spin current and the corresponding ISHE-induced charge current from $z'_{\rm Pt}=0$ up to $z_{\rm Pt}$ in Pt. The resulting effective SHA, $\Theta_{\rm eff}$, is plotted in the inset to \cref{fig12} as a function of $z_{\rm Pt}$. Using \eqref{eq:thetaeff}, we fit $\Theta_{\rm eff}$ to obtain $\Theta_{\rm I}=35 \pm 10 \%$, almost 10 times larger than the bulk SHA of Pt. This estimate of $\Theta_{\rm I}$ only includes the interface contribution on the Pt side but misses the negative values on the Au side shown in \cref{fig12}. To account for the whole interface, the formulation of \cref{subsec:ishaint} is generalized in \Cref{sec:app} where an improved estimate of $\Theta_{\rm Au|Pt}=19\pm 6\%$ is found.

\subsubsection{Three center terms}
\label{subsec:3CT}

\begin{figure}[t]
\includegraphics[width=8.6cm]{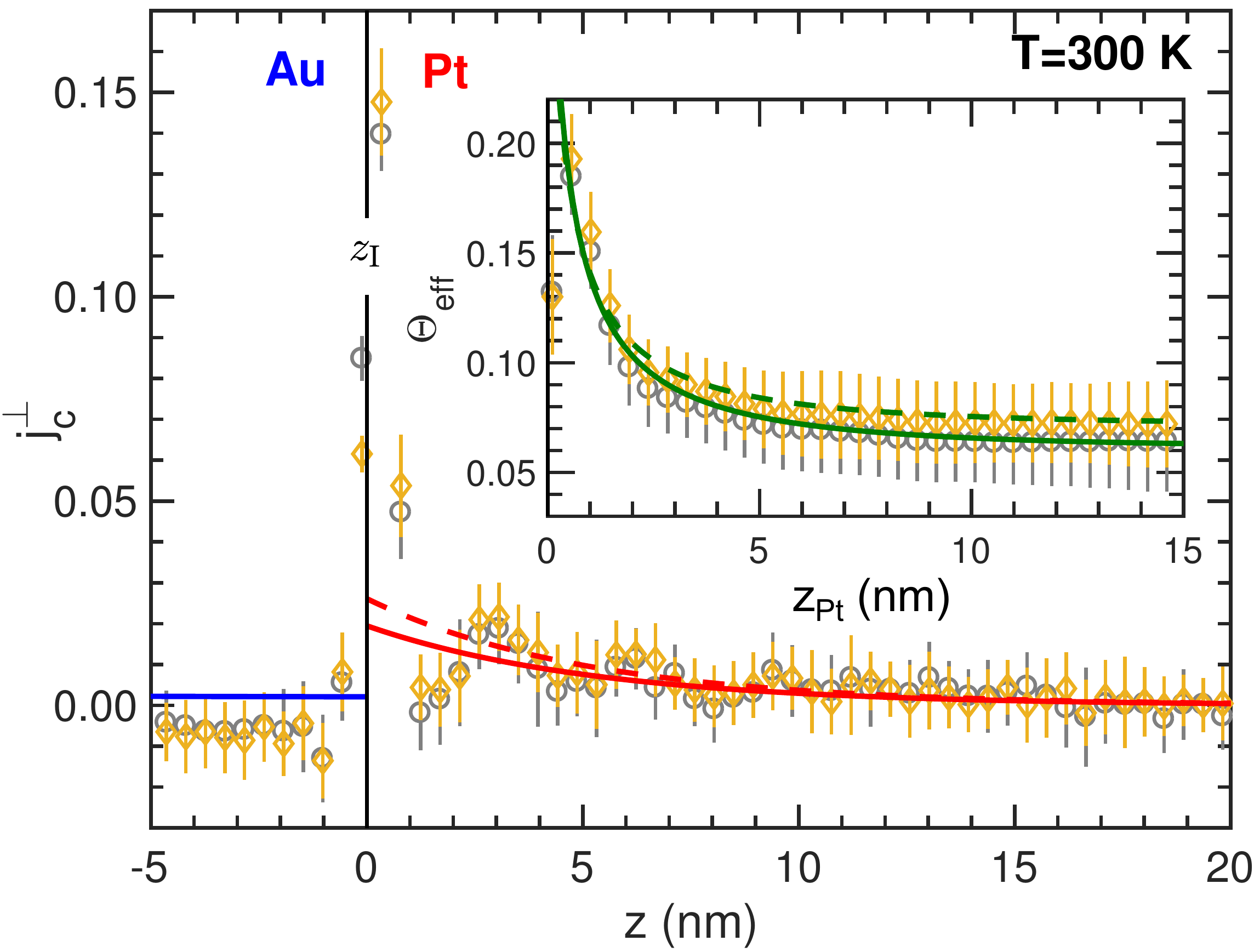}
\caption{Comparison of the ISHE induced transverse charge current $j_c^{\perp}$ in the $y$ direction calculated with two-center terms (grey circles) and three-center terms (yellow diamonds) for a Au(10 nm)$|$Pt(30 nm) bilayer embedded between ballistic Au leads at 300 K. The red lines indicate the fitted $j_{s,{\rm Pt}}(z)$ from \cref{fig9} multiplied by the bulk SHAs $\Theta_{\rm Pt}=3.7\%$ (two center terms, solid line) and $\Theta_{\rm Pt}=5.0\%$ (three-centre terms, dashed line). Inset: Effective SHA $\Theta_{\rm eff}$ calculated using two and three center terms, in grey circles and yellow diamonds, respectively. The solid and dashed green lines indicate the corresponding fits obtained using \eqref{eq:thetaeff}.}
\label{fig13}
\end{figure}

In previous work \cite{Wesselink:prb19}, we compared $l_{\rm Pt}$ and $\Theta_{\rm Pt}$ obtained using two- and three-center terms in the SOC part of the Hamiltonian. On including three-center terms, $l_{\rm Pt}$ decreased by 5\% from 5.25 to 4.96 nm while $\Theta_{\rm Pt}$ increased by 35\%, from 3.7\% to 5.0\%. This sensitivity is related to the Fermi level being close to a peak in the density of states that makes Fermi surface properties very sensitive to details of the SOC implementation. The peak corresponds to a van Hove singularity and is a consequence of the three dimensional translational symmetry of bulk Pt. For the present Au$|$Pt bilayer, we find that the spin current, and thus the SML, do not depend on the three-center terms.
In \cref{fig13}, we compare the ISHE induced charge current for the two cases. A slight difference in $j_c^{\perp}$ in the vicinity of the interface is visible. By plotting the effective SHA in the inset to \cref{fig13}, we observe that $\Theta_{\rm eff}$ calculated with three-center terms (yellow diamonds) appears simply shifted compared to the original data (grey circles) by virtue of the higher value of $\Theta_{\rm Pt}$. By fitting to \eqref{eq:thetaeff}, we find $\Theta_{\rm I}=33 \pm 11\%$ compared to $\Theta_{\rm I}=35 \pm 10\%$ obtained using only two-center terms. Thus $\Theta_{\rm I}$ is not affected by the choice of two- or three-center terms within the accuracy of the calculations.

\subsubsection{Lattice mismatch}
\label{sssubsec:lm2}
If ISHE calculations are made at 300 K for a (111) Au$|$Pt interface with both Au and Pt at their equilibrium bulk volumes as described in \cref{sssubsec:lm}, we find that $\Theta_{\rm I}$ increases to $46 \pm 18\%$. The large error bar makes it impossible to decide whether this increase is significant. It would, however, be in line with the observed trend for the other interface parameters and we would not be surprised to see an increase in $\Theta_{\rm I}$ for a calculation with smaller errors.

\subsection{Au$|$Pt vs Au$|$Pd}
\label{subsec:AuPt_AuPd}

\begin{figure}[t]
\includegraphics[width=8.6cm]{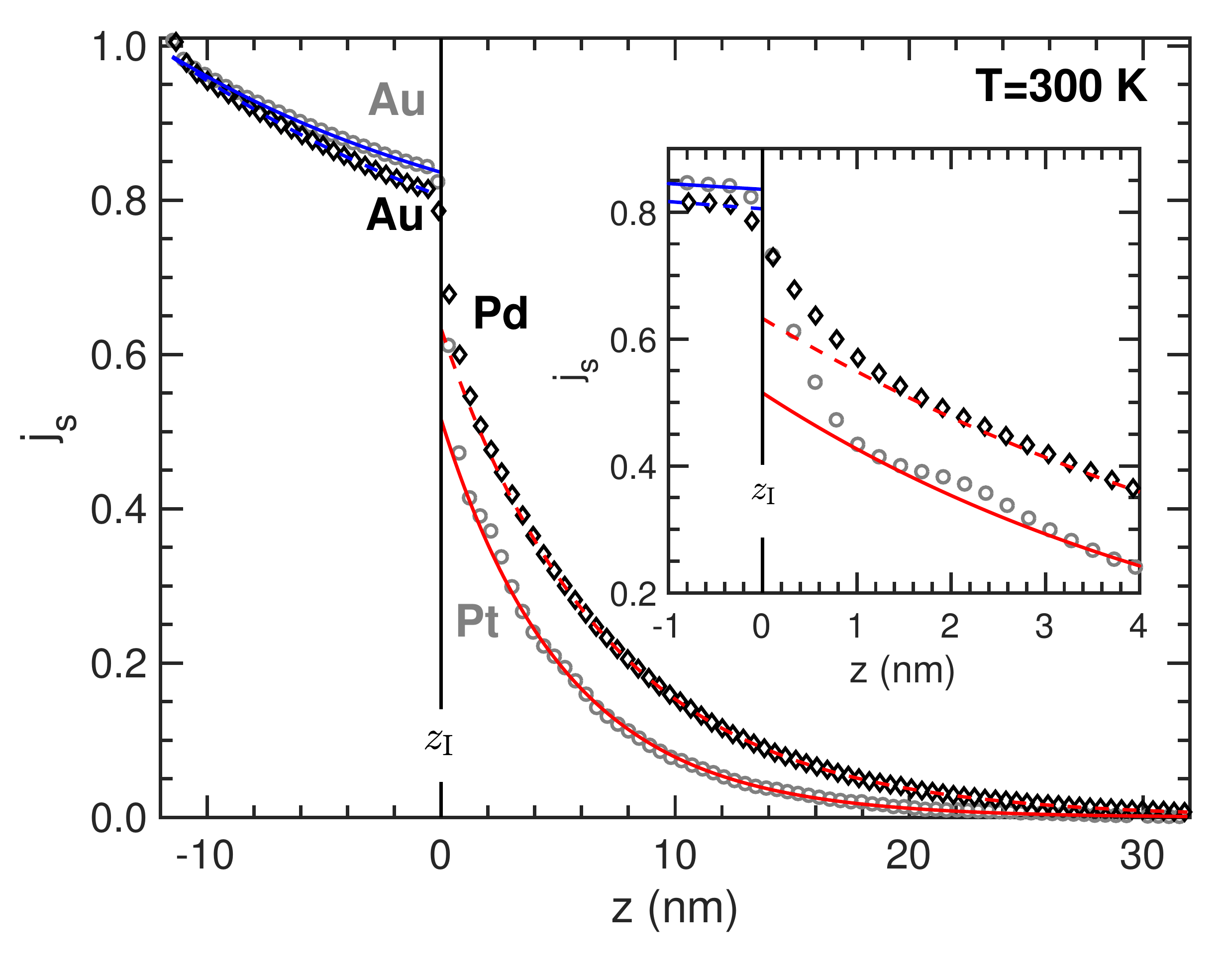}
\caption{Spin-current density calculated for a Au(10 nm)$|$Pd(30 nm) bilayer (black diamonds) sandwiched between ballistic Au leads at 300 K compared to the corresponding results for a Au(10)$|$Pt(30) bilayer (grey circles) from \cref{fig9}; in both cases the lattice parameter of Au is chosen to be that of the other metal. The solid and dashed lines indicate the fits for $j_s$ in individual layers using \eqref{eq:jsi3} for Au$|$Pt and Au$|$Pd bilayers respectively. Inset: a closer look at the rapid decay of the spin current in the vicinity of the interfaces.}
\label{fig14}
\end{figure}

Now that we have described how $AR_{\rm I}$, $\delta$ and $\Theta_{\rm I}$ are determined for Au$|$Pt, we repeat the procedure for the Au$|$Pd interface and compare the results for the two. We begin by extracting the bulk parameters for Pd that are needed for the fits that will result in the interface parameters. We find $l_{\rm Pd}=7.06 \pm 0.02$~nm and $\Theta_{\rm Pd}=3.5 \pm 0.1 \%$ when we choose the thermal disorder to reproduce the experimental resistivity of Pd, $\rho_{\rm Pd}=10.8~\mu\Omega\,\rm cm$ at 300 K \cite{HCP90}.

In \cref{fig14}, we compare the effect of spin-flip diffusion in Au$|$Pt (grey circles) and Au$|$Pd (black diamonds) bilayers at room temperature. Small differences are visible in the decrease of $j_s(z)$ in Au for the two systems. One reason is that the interface reflectivity determining the coefficient $A_i$ of the increasing exponential term in \eqref{eq:jsi3} is different, as we will see below. In addition, the lattice constant of Au is matched to that of Pt or Pd so is not the same for the two bilayers. In the vicinity of the interface, the rapid decay of $j_s(z)$ is more prominent and much sharper in Pt than in Pd, as shown in the inset to \cref{fig14}. By fitting $j_s(z)$ to \eqref{eq:jsi3}, we obtain $j_{s,\rm Au}(z_{\rm I})$ and $j_{s,\rm Pd}(z_{\rm I})$ by extrapolation to the interface. A value of $AR_{\rm Au|Pd}=0.36\pm0.04~{\rm f}\Omega\,{\rm m^2}$ is obtained using the procedure described in \cref{subsec:intRAuPt}; this is smaller than the corresponding value of $0.54\pm0.03~{\rm f}\Omega\,{\rm m^2}$ we found for Au$|$Pt. By substituting all the input parameters and their uncertainities into \eqref{eq:delta}, a value of $\delta_{\rm Au|Pd}=0.32\pm0.02$ is extracted numerically, which is approximately half of what we found for Au$|$Pt, $\delta_{\rm Au|Pt}=0.62\pm0.03$ \footnote{An error was made in the calculation of $AR_{\rm Au|Pd}$ in \cite{Gupta:prl20} which influenced the value of $\delta$ calculated for Au$|$Pd. Corrected values of both parameters are given in the present paper.}. 

\begin{table}[b]
\caption{Room temperature (T=300 K) transport parameters. 
(Upper) Bulk parameters: resistivity $\rho$ $(\mu\Omega \,$cm), spin-flip diffusion length $l_{\rm sf}$ (nm) and spin-Hall angle $\Theta_{\rm sH}$ (\%) for NM=Pt or Pd. 
(Lower) Interface parameters: 
interface resistance $AR_{\rm Au|NM}({\rm f}\Omega \,{\rm m}^2)$, 
spin-memory loss $\delta$ (dimensionless) and 
interface spin-Hall angle $\Theta_{\rm Au|NM}$ $(\%)$ for Au$|$Pd and Au$|$Pt interfaces. Two interfaces are considered: a pseudomorphic interface for which $a_{\rm Au}$ is chosen to be equal to $a_{\rm NM}$ (Compressed) and an interface between equilibrium Au and NM (Relaxed).  
}
\begin{ruledtabular}
\begin{tabular}{c|cccc}
Bulk NM              & \multicolumn{2}{c}{Pd}            & \multicolumn{2}{c}{Pt}\\
\hline
$\rho$               & \multicolumn{2}{c}{$10.8\pm0.1$}  & \multicolumn{2}{c}{$10.7\pm0.2$} \\  
$l_{\rm sf}$         & \multicolumn{2}{c}{$7.06\pm0.02$} & \multicolumn{2}{c}{$5.25\pm0.05$} \\ 
$\Theta_{\rm sH}$    & \multicolumn{2}{c}{$3.5\pm0.1$}   & \multicolumn{2}{c}{$3.7\pm0.1$} \\
\hline
Au$|$NM              & Compressed     & Relaxed \cite{LiuRX:prb22}          
                                                        &  Compressed   & Relaxed \\
Interface            & $a_{\rm Au} = a_{\rm Pd}$   
                                      & $a_{\rm Au} = a_{\rm Au}$   
                                                        &  $a_{\rm Au}=a_{\rm Pt}$
                                                                        & $a_{\rm Au}=a_{\rm Au}$ \\
\hline
$AR_{\rm I}$         & $0.36\pm0.04$  & $0.55\pm0.02$   & $0.54\pm0.03$ & $0.81\pm0.04$ \\     
$\delta$             & $0.32\pm0.02$  & $0.63 \pm 0.02$ & $0.62\pm0.03$ & $0.81\pm0.05$ \\
$\Theta_{\rm I}$ & $21\pm 3$          & $17\pm 6$       & $35\pm10$     & $46\pm18$     \\
\end{tabular}
\end{ruledtabular}
\label{tab:PtPd}  
\end{table}

\begin{figure}[t]
\includegraphics[width=8.6cm]{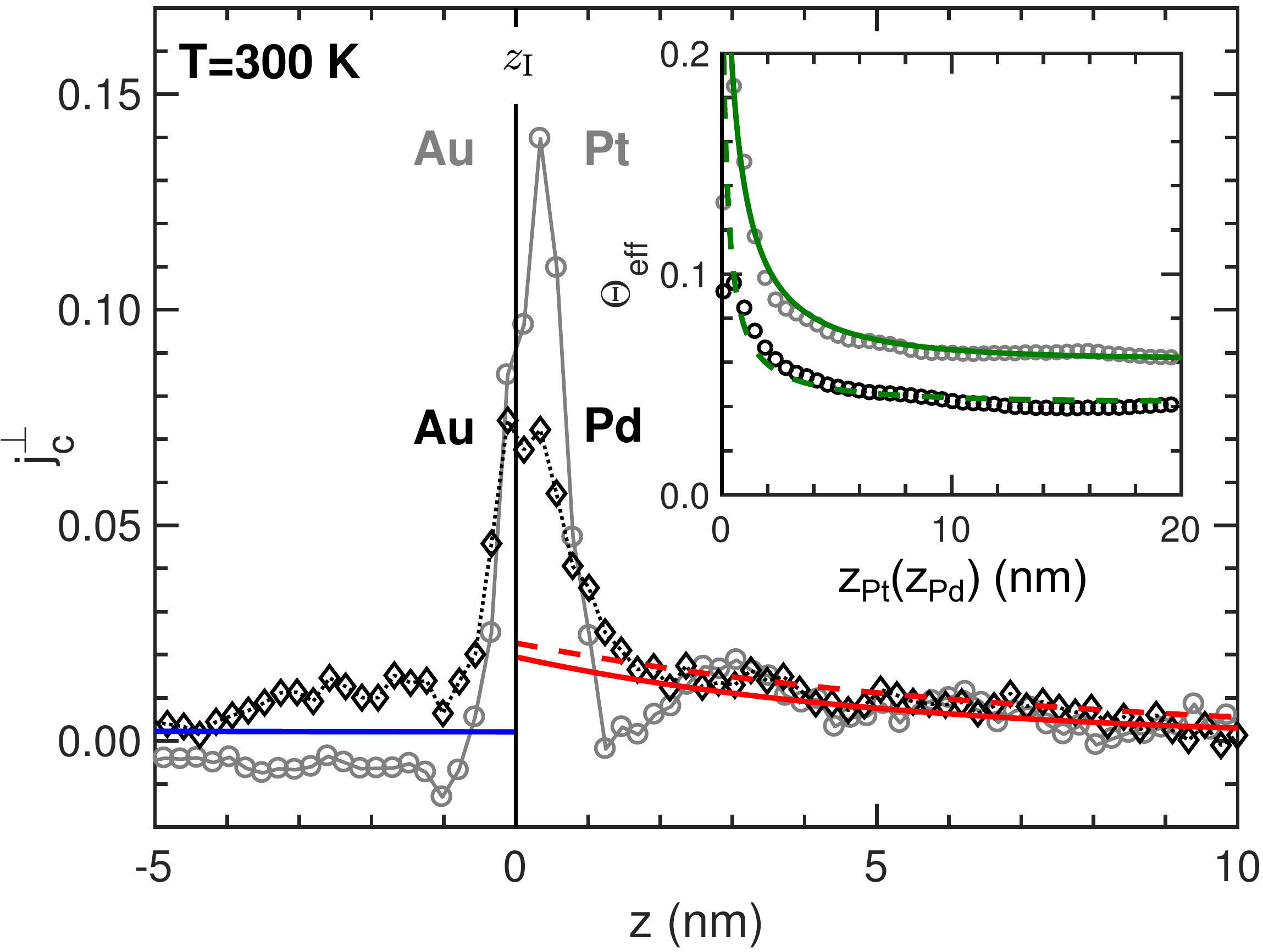}
\caption{Transverse charge current induced in the $y$ direction by the ISHE for a Au(10 nm)$|$Pt(30 nm) bilayer (grey circles) compared to that for a Au(10 nm)$|$Pd(30 nm) bilayer (black diamonds) embedded between ballistic Au leads at 300 K. Inset: Effective SHA $\Theta_{\rm eff}$ calculated for Au$|$Pt (grey circles) and Au$|$Pd (black diamonds) interfaces. The solid and dashed green lines indicate the fits obtained for Pt and Pd, respectively, using \eqref{eq:thetaeff}.
}
\label{fig15}
\end{figure}

In \cref{fig15}, we compare the ISHE-induced charge current $j_c^{\perp}(z)$ for Au$|$Pt (grey circles) and Au$|$Pd (black diamonds) at T=300~K. The peak around $z_{\rm I}$ coming from the interface ISHE described by $\Theta_{\rm I}$ is significantly lower for Au$|$Pd than for Au$|$Pt. By fitting $\Theta_{\rm eff}$, we find $\Theta_{\rm I}=21 \pm 3 \%$ for Au$|$Pd. There are a few other interesting features. At a distance greater than 3 nm from the interface, $j_c^{\perp}(z)$ in Pd and Pt appear almost identical. Given that we find $\Theta_{\rm Pd}=3.5\%$ is only 5\% smaller than $\Theta_{\rm Pt}=3.7 \%$, this is not surprising. On the other hand, $j_c^{\perp}$ in Au gradually decreases towards a small positive value away from the Au$|$Pd interface in contrast to the small negative value we see for Au in the Au$|$Pt bilayer. This is because the smaller SOC of Pd compared to thatof Pt does not induce a significant filtering effect by the Rashba-type spin-orbit interaction which we used to explained the negative effective SHA in Au close to the Au$|$Pt interface. It highlights that the magnitude of the Rashba effect at the Au$|$Pt interface is  mainly determined by Pt.



For fully relaxed Au$|$Pt and Au$|$Pd geometries, the values of $AR_{\rm I}$, $\delta$ and $\Theta_{\rm I}$ calculated at 300 K  are compared with the results obtained by matching the Au lattice parameter to those of Pd or Pt in \cref{tab:PtPd}. Lattice mismatch is seen to increase both the interface resistance $AR_{\rm I}$ and the SML parameter $\delta$ substantially for both Pd and Pt. In particular, that the value of $\delta_{\rm Au|Pt}=0.81\pm0.05$ obtained for the relaxed Au$|$Pt interface is larger than $\delta_{\rm Au|Pd}=0.63\pm0.02$ is attributed to the SOC of Pt being larger than that of Pd. In fact, this value $\delta_{\rm Au|Pt}=0.81\pm0.05$ is comparable with the SML calculated for the Cu$|$Pt interface $\delta_{\rm Cu|Pt}=0.77\pm0.04$ \cite{LiuRX:prb22} suggesting that the free-electron-like conduction electrons in Au do not play a key role in interface dissipation of spin currents. The larger SOC in Pt also leads to a larger $\Theta_{\rm I}=46 \pm 18 \%$ for Au$|$Pt compared to $\Theta_{\rm I}=17 \pm 6 \%$ for Au$|$Pd for relaxed interfaces, \cref{tab:PtPd}. Because of the large uncertainty in the calculated values of $\Theta_{\rm I}$, we cannot draw strong conclusions about the role of lattice mismatch on the interface SHA.

\subsection{Temperature dependence of the interface parameters}
\label{subsec:temp}

\begin{figure}[t]
\includegraphics[width=8.6cm]{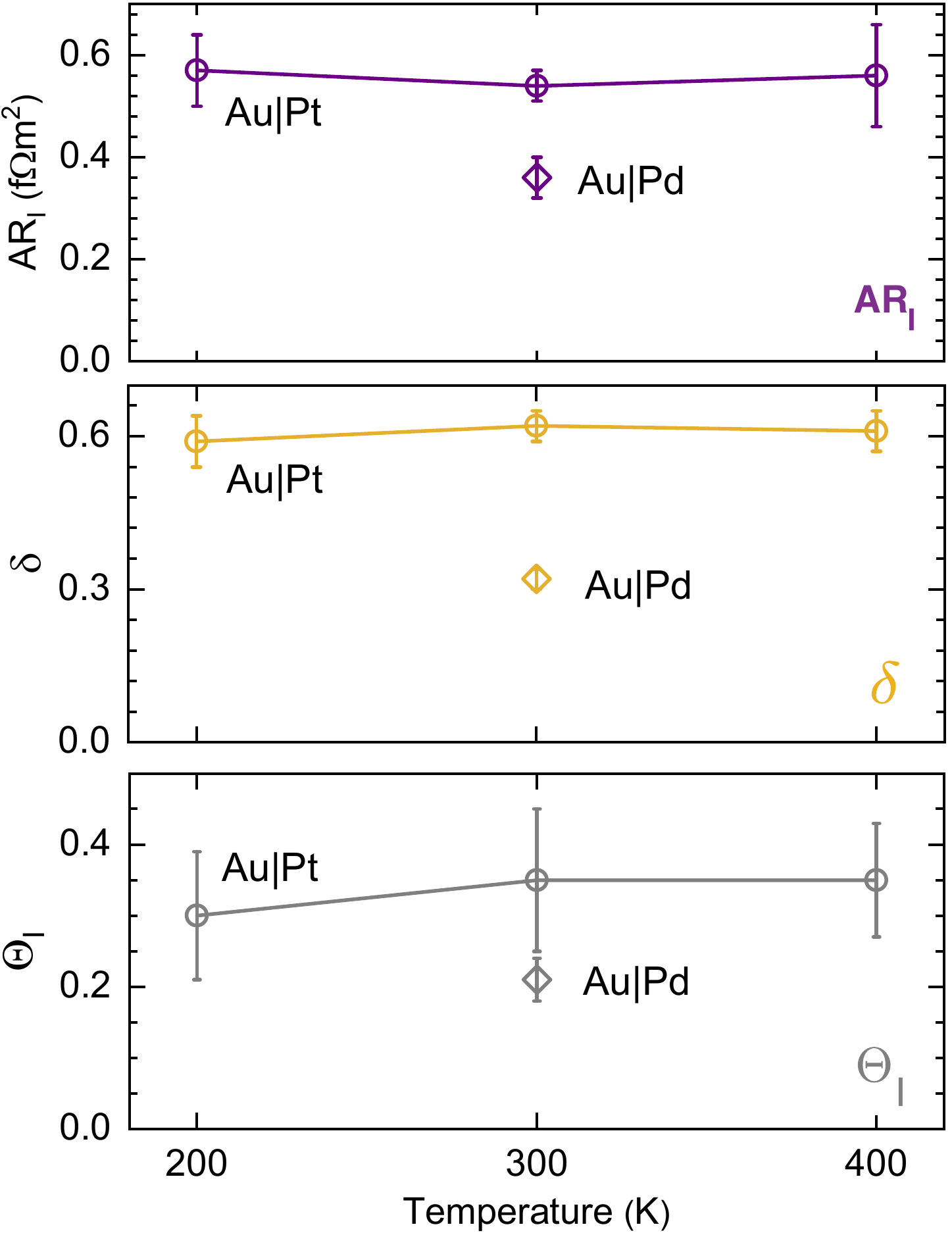}
\caption{Temperature dependence of the interface parameters $AR_{\rm I}$ (purple), $\delta$ (yellow) and $\Theta_{\rm I}$ (grey) for a Au$|$Pt interface (circles). The corresponding set of parameters for Au$|$Pd at 300 K are included (diamonds).
}
\label{fig16}
\end{figure}

While the bulk parameters $\rho_{\rm Pt}$, $1/l_{\rm Pt}$ and $\Theta_{\rm Pt}$ increase linearly with temperature \cite{HCP90, LiuY:prb15, Isasa:prb15a, WangL:prl16}, virtually nothing is known about the temperature dependence of interface parameters. We therefore determined the Au$|$Pt interface parameters at 200 and 400 K and plot them in \cref{fig16} together with the 300 K results already calculated. Within the error bars of the calculations, all three parameters $R_{\rm I}$, $\delta$ and $\Theta_{\rm I}$ are constant between 200 and 400 K, in contrast to what happens when the interface involves a ferromagnetic material \cite{Gupta:prl20}. 

\section{Discussion and Conclusions}
\label{sec:Conclusions}

In this paper, we have described a scheme to extract $AR_{\rm I}$ and $\delta$ for nonmagnetic Au$|$Pt and Au$|$Pd interfaces at finite temperatures from local spin and charge currents \cite{Wesselink:prb19} calculated from the results of first principles fully relativistic scattering calculations \cite{Starikov:prl10, *Starikov:prb18}. We also extracted the interface contribution to the SHE characterized by a dimensionless SHA $\Theta_{\rm I}$ \cite{WangL:prl16}. Table \ref{tab:PtPd} summarizes the results for the Pd and Pt bulk parameters as well as the interface parameters for the corresponding interfaces with Au at T=300~K. We found a substantial spin-memory loss and an interface SHA that is almost an order of magnitude larger than the bulk SHA for both interfaces.

By studying the effect of the intermixing of interface layers, modelled as a $\rm Au_{50}Pt_{50}$ alloy, as well as lattice mismatch for a Au$|$Pt interface at 300 K, we found that incommensurate interfaces significantly enhance the interface parameters because of the increased momentum-nonconserving scattering suggesting directions to be explored experimentally to attain smaller values of these interface parameters with cleaner and lattice matched interfaces.

Because of their relatively weak effective SOC, the free electron like metals Cu, Ag and Au are often used as spacer layers in spin-pumping and related experiments to suppress the magnetization induced in Pd and Pt by proximity to a magnetic layer, the so-called ``magnetization proximity effect'' (MPE); this is considered to have a significant influence on spin transport through interfaces \cite{Huang:prl12, Nan:prb15}. While the role of the MPE and whether a spacer layer modifies the interface effects are still being debated \cite{Weiler:prl13, Boone:jap15, Rojas-Sanchez:prl14, ZhangW:natp15, Zhu:prb18}, our findings show that when nonmagnetic spacer layers are introduced, additional interface parameters must also be introduced to describe the spin memory loss and interface SHE at the new interfaces, even when these are between nonmagnetic materials.
Many experiments use materials like Ta as capping or seeding layers adjacent to Pt \cite{Nguyen:prl16, Berger:prb18b}. For small thicknesses of Pt, an interface with Ta may also lead to an enhanced interface SHE and spin memory loss. Not taking this into consideration will most likely influence the values of ``bulk'' parameters extracted for Pt.

Experimental \cite{Jungfleisch:prb16} and theoretical \cite{Amin:prl18} studies have shown that nonmagnetic interfaces can generate spin currents and exert torques on neighbouring magnetic layers. This again points towards the importance of interface spin-orbit splitting for nonmagnetic interfaces and the large values of the interface SHA we observe support these studies. The interface SHA could be employed as a parameter that determines the efficiency of these spin currents.

The stronger SOC in Pt compared to Pd leads to a larger $\delta$ and $\Theta_{\rm I}$ for Au$|$Pt compared to Au$|$Pd. It is important to note that the bulk SHAs that we find for the two materials differ by only 5\%, although numerous studies suggest a larger SHA for Pt compared to Pd \cite{Sinova:rmp15}. In addition to identifying the source of the wide  spread in values reported for the bulk SHA, clarifying the role of the SML and the interface SHA is essential if reliable and reproducible values of the SHA characteristic of bulk Pd and Pt are to be determined.

Schep {\em et al.} developed a model for the resistance $AR_{\rm A|B}$ of an A$|$B interface in terms of the transmission through the interface between the ballistic (T=0~K) materials; the interface is then embedded between diffusively scattering A and B materials \cite{Schep:prb97, Schep:jmmm98} and it is implicitly assumed that the interface resistance does not depend on temperature. The temperature-independent behaviour we found for $AR_{\rm Au|Pt}$ in \cref{subsec:temp} is consistent with this. We can take a further step and calculate $AR_{\rm Au|Pt}$ using Schep's ansatz, both with and without SOC. We do so for the lattice matched (compressed Au) case. Compared to the room temperature value of $0.54\pm0.03\,$f$\Omega \,{\rm m}^2$ calculated from the currents, we find $AR_{\rm Au|Pt}=0.63\,$f$\Omega\, {\rm m}^2$ with SOC and $0.56\,$f$\Omega\, {\rm m}^2$ without SOC using Schep's procedure. The good agreement confirms that Schep's ansatz describes the essence of the problem. 

Our results for $AR_{\rm I}$ and $\delta$ for lattice-matched Au$|$Pd (111) interfaces can be compared to those estimated theoretically by Flores {\em et al.} \cite{Flores:prb20} who combined Schep's ansatz with circuit theory and first-principles calculations of the scattering matrix to determine $AR_{\rm I}$ and $\delta$ for numerous interfaces (but not for Au$|$Pt). For a clean Au$|$Pd interface they report values of $AR_{\rm Au|Pd}$ between 0.83 and 0.87 f$\Omega\,\rm m^2$ (depending on the spin-orientation); these should be compared to our corresponding value of $AR_{\rm Au|Pd}=0.36\pm0.04~{\rm f}\Omega\,{\rm m^2}$ and a low-temperature experimental value  of $AR_{\rm Au|Pd}=0.23\pm0.08~{\rm f}\Omega\,{\rm m^2}$ \cite{Galinon:apl05, Bass:jmmm16}. With 50\%-50\% intermixing in two interface layers, their $AR_{\rm Au|Pd}$ increases to  between 0.95 and 0.99 f$\Omega\,\rm m^2$ while we find a value of $AR_{\rm Au|Pt}=0.76\pm0.04~{\rm f}\Omega\,{\rm m^2}$ for Au$|$Pt and expect the value for Au$|$Pd to be lower. For $\delta$, Flores reports values of $\delta$ between 0.53 and 0.73 (depending on the spin orientation) for a clean interface increasing to between 0.58 and 0.82 with two intermixed interface layers. Our corresponding value for $\delta$ is $0.32 \pm 0.02$ for a clean interface that we expect to increase substantially with intermixing by analogy with Au$|$Pt. The advantage of our approach is that temperature is taken into account explicitly and we show that is has little effect for the parameters describing transport through nonmagnetic interfaces. By fitting our results for the spin currents with the same phenomenological theory used to interpret experiment, there is a one-to-one correspondence between experimental and theoretical parameters.

\acknowledgements{
This work was financially supported by the ``Nederlandse Organisatie voor Wetenschappelijk Onderzoek'' (NWO) through the research programme of the former ``Stichting voor Fundamenteel Onderzoek der Materie,'' (NWO-I, formerly FOM) and through the use of supercomputer facilities of NWO ``Exacte Wetenschappen'' (Physical Sciences). K.G. acknowledges funding from the Shell-NWO/FOM “Computational Sciences for Energy Research” PhD program (CSER-PhD; nr. i32; project number 13CSER059). The work was also supported by the Royal Netherlands Academy of Arts and Sciences (KNAW). Work at Beijing Normal University was supported by National Natural Science Foundation of China (Grant No. 12174028).}
 
\appendix
\section{Extracting $\Theta_{\rm I}$ including contributions from both sides of the NM$|$NM$'$ interface}
\label{sec:app}

In the formalism presented in \cref{subsec:ishaint}, the interface SHA $\Theta_{\rm I}$ is extracted by integrating from the interface $z_{\rm I}=0$ to a  distance $z=L_{\rm NM'}$ to the right of the NM$|$NM$'$ interface so only the contribution from the NM$'$ side of the interface is included. In \cref{fig12} we see that the negative Au contribution from near the Au$|$Pt interface may reduce the total $\Theta_{\rm I}$ but the method described in \cref{subsec:ishaint} does not allow us to quantitatively evaluate this contribution. In this Appendix,  the theoretical formalism of \cref{subsec:ishaint} is generalized to include contributions from both sides of the NM$|$NM$'$ interface in $\Theta_{\rm I}$. 

The longitudinal spin current is decomposed into bulk and interface contributions as 
\begin{eqnarray}
 j_{sx}^z(z)&=& \left( E_1 e^{-z/l_1}-F_1 e^{z/l_1} \right)  \theta(-z)  \nonumber\\
 			&&+ \bar{J}_s^{\rm I} \delta(z) 
 			+ E_2 e^{-z/l_2} \theta(z),
\label{eq:AA1}
\end{eqnarray}
where $E_i=\frac{B_i}{2ej\rho_i l_i}$ ($i=1,2$) and $F_1=\frac{A_1}{2ej\rho_1 l_1}$ with $A_i$ and $B_i$ defined in \eqref{eq:jsi3}. $\bar{J}_s^{\rm I}$ is the effective spin current density at the interface and $\theta(z)$ is the unit step function.
 
To interpret the ISHE for a Au$|$Pt bilayer and obtain a quantitative value of the interface SHA $\Theta_{\rm I} \equiv \Theta^{\rm I}_{\rm sH}$, we consider the Au$|$Pt segment extending from a position $z=-L$ through the interface at $z_{\rm I}=0$ to a position $z=+L$ where the total transverse electron current 
\begin{equation}
\bar J_c(L) \equiv \int_{-L}^{+L} j^y_c(z)dz 
\label{eq:AA2}
\end{equation}
is generated by the total spin current 
\begin{equation}
\bar J_s(L) \equiv \int_{-L}^{+L} j^z_{sx}(z)dz
\label{eq:AA3} 
\end{equation}
obtained by integrating \eqref{eq:AA1} so $\bar J_s(L)$ can be written as 
\begin{widetext}
\begin{equation}
\bar J_s(L)= E_1 l_1 \left( e^{+L/l_1} - 1 \right)+F_1 l_1 \left( e^{-L/l_1} - 1 \right)
 			+ \bar{J}_s^I
 			+ E_2 l_2 \left( 1-e^{-L/l_2} \right).
\label{eq:AA4}        
\end{equation}
The total transverse charge current density can be calculated with the interface and bulk SHAs as
\begin{eqnarray}
\bar{J}_c&=& \int_{-L}^{+L}
\left[ \Theta_1 \left( E_1 e^{-z/l_1}-F_1 e^{z/l_1} \right)
 		+ \Theta_{\rm I} \bar{J}_s^I\delta(z)
 		+ \Theta_2 E_2 e^{-z/l_2} \right] dz   \nonumber\\
 		&=& \Theta_1 \left[ E_1 l_1 \left( e^{+L/l_1} - 1 \right)+F_1 l_1 \left( e^{-L/l_1} - 1 \right) \right]
 		+ \Theta_{\rm I} \bar{J}_s^{\rm I} 
 		+ \Theta_2 E_2 l_2 \left( 1-e^{-L/l_2} \right).
\label{eq:AA5} 
\end{eqnarray}
Here the spin-flip diffusion length $l_i$ and the bulk SHA $\Theta_i$ can be determined in separate calculations for bulk material NM$_i$.
If we interpret the interface spin-Hall contribution in terms of an effective value $\Theta_{\rm eff} \equiv \bar J_c/\bar J_s$, then
\begin{equation}
\Theta_{\rm eff}(L)= \frac{\bar J_c(L)}{\bar J_s(L)}
 		= \frac{\Theta_1 \big[E_1 l_1 \left( e^{+L/l_1} -1 \right)+F_1 l_1 \left( e^{-L/l_1} -1 \right)\big]
 		+ \Theta_{\rm I} \bar{J}_s^{\rm I} 
 		+ \Theta_2 E_2 l_2 \left( 1-e^{-L/l_2} \right)}
 		{E_1 l_1 \left( e^{+L/l_1} -1 \right)+F_1 l_1 \left( e^{-L/l_1} -1 \right)
 		+ \bar{J}_s^I +
 		+ F_2 l_2 \left( 1-e^{-L/l_2} \right)}.
\label{eq:AA6} 
\end{equation}
\end{widetext}

\begin{figure}[t]
\includegraphics[width=8.6cm]{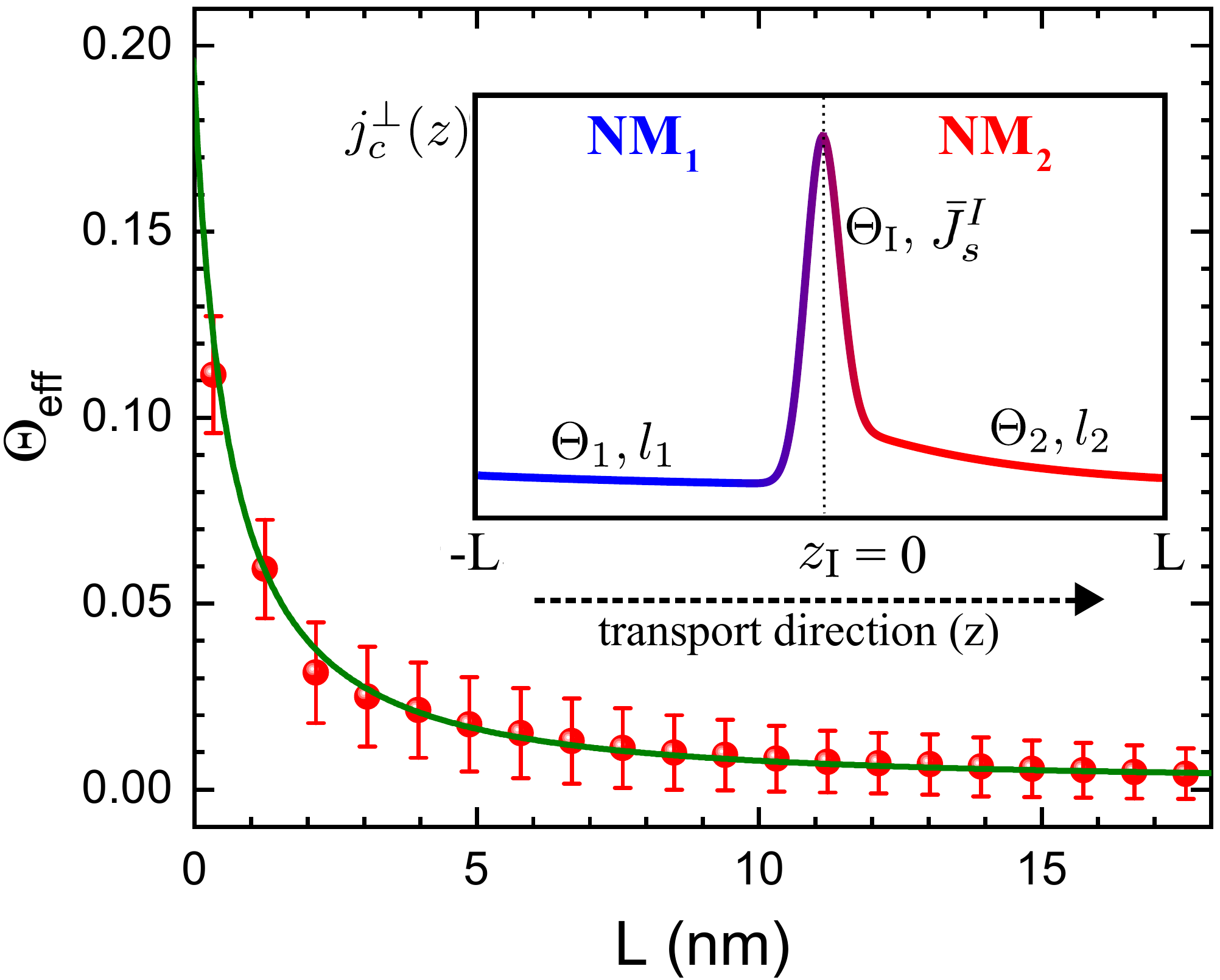}
\caption{Effective SHA calculated by integrating the corresponding transverse charge current density and longitudinal spin current density. The solid green line is a fit using \eqref{eq:AA6}. }
\label{figappendix}
\end{figure}

$\Theta_{\rm eff}(L)$ can be determined from \eqref{eq:AA2} and \eqref{eq:AA3} and fitted to the analytical form \eqref{eq:AA6} where everything is known except $\Theta_{\rm I}$. Unlike in Wang {\em et al.} \cite{WangL:prl16}, both $\bar J_c(L)$ and $\bar J_s(L)$ increase with $L$ and the contributions from the Au and Pt sides of the interface are included naturally. The calculated $\Theta_{\rm eff}(L)$ for the commensurate Au$|$Pt interface is plotted as a function of $L$ in \cref{figappendix} as red dots. Taking the bulk SHAs, $\Theta_{\rm Au}=0.25\%$, $\Theta_{\rm Pt}=3.7\%$, the spin-flip diffusion length of Au, $l_{\rm Au}=50.9$~nm and of Pt, $l_{\rm Pt}=5.25$~nm, we are able to fit the calculated $\Theta_{\rm eff}$ using Eq.~(\ref{eq:AA6}). The fit illustrated by the solid green line describes the calculated data points perfectly. The value we obtain for the interface SHA, $\Theta_{\rm I}=19\pm 6\%$. It is smaller than the value $\Theta_{\rm I}=35\pm 10\%$ we obtained in \cref{subsec:thetaI} considering the contribution on the Pt side only. This is because of the negative contribution on the Au side, which extends further into Au than 10 nm. Even in the limit of thick Pt, $\Theta_{\rm eff}$ does not approach the bulk value $\Theta_{\rm Pt}=3.7\%$ indicating that it is essential to explicitly include an interface contribution.

As shown in \cref{fig11disorder} for Au$|$Pt with interface mixing, it can be important to include the contributions from both sides of the interface to estimate $\Theta_{\rm I}$ quantitatively.  Using \eqref{eq:AA6}, we find $\Theta_{\rm I}=37\pm8\%$ and $31\pm4\%$ for 2 layers and 4 layers of Au$_{50}$Pt$_{50}$ interface alloy, respectively. These values are nearly twice the value ($19\pm 6\%$) obtained for the clean interface. 

%


\end{document}